\documentclass[letterpaper,twocolumn,10pt]{article}
\PassOptionsToPackage{table}{xcolor}
\usepackage{usenix}
\usepackage{amssymb}
\usepackage{subcaption}

\usepackage{microtype}
\usepackage{booktabs}
\usepackage{csquotes}
\usepackage{xspace}
\usepackage{acronym}
\usepackage{amsmath}
\usepackage[indLines]{algpseudocodex}

\usepackage{tcolorbox}
\tcbuselibrary{skins,breakable}
\usepackage{listings}
\definecolor{railgunCodeBg}{RGB}{241,241,237}
\definecolor{railgunCodeRule}{RGB}{80,80,80}
\definecolor{railgunCodeComment}{RGB}{0,120,0}
\definecolor{railgunCodeKeyword}{RGB}{0,0,210}
\definecolor{railgunCodeString}{RGB}{163,21,21}
\definecolor{railgunCodeType}{RGB}{111,0,138}
\definecolor{railgunCodeFunction}{RGB}{0,105,120}
\lstdefinelanguage{ArtifactRust}{
  sensitive=true,
  morekeywords={as,async,await,break,const,continue,crate,dyn,else,enum,extern,false,fn,for,if,impl,in,let,loop,match,mod,move,mut,pub,ref,return,self,Self,static,struct,super,trait,true,type,unsafe,use,where,while},
  morekeywords=[2]{Err,None,Ok,Some,String,Vec},
  morekeywords=[3]{clone,cloned,contains_key,copied,format,get,insert,into,is_empty,ok_or_else,push},
  morecomment=[l]{//},
  morecomment=[s]{/*}{*/},
  morestring=[b]"
}
\lstdefinelanguage{ArtifactJavaScript}{
  sensitive=true,
  morekeywords={async,await,break,case,catch,class,const,continue,debugger,default,delete,do,else,export,extends,false,finally,for,function,if,import,in,instanceof,let,new,null,of,return,super,switch,this,throw,true,try,typeof,var,void,while,with,yield},
  morekeywords=[2]{Array,Map,Number,Set,String},
  morekeywords=[3]{compareArrays,deriveMetrics,digestOrdinals,flatMap,forEach,get,ledgerPosition,map,precedes,push,requireCondition},
  morecomment=[l]{//},
  morecomment=[s]{/*}{*/},
  morestring=[b]",
  morestring=[b]'
}
\lstdefinestyle{artifactpatch}{
  basicstyle=\ttfamily\scriptsize\color{black},
  numbers=left,
  numberstyle=\tiny\color{gray},
  numbersep=7pt,
  backgroundcolor=\color{railgunCodeBg},
  frame=tb,
  framerule=0.7pt,
  framesep=5pt,
  xleftmargin=1.1em,
  xrightmargin=0.2em,
  rulecolor=\color{railgunCodeRule},
  breaklines=true,
  columns=fullflexible,
  keepspaces=true,
  showstringspaces=false,
  aboveskip=2pt,
  belowskip=2pt,
  keywordstyle=\color{railgunCodeKeyword},
  keywordstyle=[2]\color{railgunCodeType},
  keywordstyle=[3]\color{railgunCodeFunction},
  commentstyle=\color{railgunCodeComment},
  identifierstyle=\color{black},
  stringstyle=\color{railgunCodeString},
}
\usepackage{varwidth}
\usepackage{multirow}
\usepackage{enumitem}
\usepackage{adjustbox}
\usepackage{pifont}
\usepackage{xurl}

\usepackage{tikz}
\DeclareRobustCommand{\fcirc}[1]{%
  \tikz[baseline=(C.base)]{
    \node[
      circle,
      fill=black,
      text=white,
      inner sep=0.1em,
      minimum size=0.1em,
      font=\bfseries\footnotesize
    ] (C) {#1};
  }%
}

\makeatletter
\@ifpackageloaded{hyperref}{}{\usepackage{hyperref}}

\makeatother
\newcommand{\appref}[1]{\hyperref[#1]{Appendix~\ref*{#1}}}
\newcommand{\point}[1]{\par\smallskip\noindent\textbf{#1.} }
\acrodef{UTXO}[UTXO]{unspent transaction output}
\acrodefplural{UTXO}[UTXOs]{unspent transaction outputs}
\acrodef{DeFi}[DeFi]{decentralized finance}
\acrodef{EVM}[EVM]{Ethereum Virtual Machine}
\acrodef{NFT}[NFT]{non-fungible token}
\acrodefplural{NFT}[NFTs]{non-fungible tokens}
\acrodef{ZKP}[ZKP]{zero-knowledge proof}
\acrodefplural{ZKP}[ZKPs]{zero-knowledge proofs}
\acrodef{SNARK}[SNARK]{succinct non-interactive argument of knowledge}
\acrodefplural{SNARK}[SNARKs]{succinct non-interactive arguments of knowledge}
\acrodef{TVL}[TVL]{total value locked}
\acrodef{OFAC}[OFAC]{Office of Foreign Assets Control}

\begin{document}

\date{}
\title{\Large \bf The Anonymity Gap: Understanding Real Privacy in Shielded \acs{UTXO}-based Protocols for \acs{DeFi}}
\author{
{\rm Hanze Guo}\\
UCL Centre for Blockchain Technologies, UK
\and
{\rm Stefanos Chaliasos}\\
UCL Centre for Blockchain Technologies, UK\\
zkSecurity, USA
\and
{\rm Yebo Feng}\\
Nanyang Technological University, Singapore
\and
{\rm Jiahua Xu}\\
UCL Centre for Blockchain Technologies, UK\\
Exponential Science, Switzerland
}
\maketitle

\begin{abstract}
    Shielded \acs{UTXO}-based protocols are becoming a core form of privacy infrastructure for \acs{DeFi}. Unlike mixers that organize privacy mainly around deposits and withdrawals, these protocols allow assets, once inside the shielded pool, to continue moving and being re-spent within the hidden state, and to become public only when users withdraw or interact with public \acs{DeFi} protocols. Their anonymity is therefore no longer a flat pool-size problem, but a provenance problem that propagates across the note/\acs{UTXO}, proof, and transaction layers. Yet, a unified analysis framework for this setting is still missing. We propose a layered system model and an analysis pipeline that uses prior history as the temporal baseline, applies cumulative pruning and cross-proof propagation to each proof's Commitment Set, and recursively traces the survivors through historical hidden-state transitions to derive the final transaction-level Anonymity Set Size.

We evaluate our methodology on the complete on-chain histories of all four Railgun production deployments and five independent Hinkal pools across six \acs{EVM} chains, analyzing 186,356 unshielding spend transactions. Using only public protocol traces and constraints, our non-heuristic analysis \emph{yields mean Anonymity Set Size reductions of 40.1\%--59.0\%} relative to each deployment's temporal baseline; \emph{3,679 transactions retain at most 10 addresses, including 1,228 singletons}. Public token constraints are the strongest and most stable source of pruning in both protocols, while the effects of tree number, proof roots, and value constraints vary with protocol design and historical state. Together with representative cases, these results reveal interpretable anonymity-loss patterns and implications for user behavior and future protocol design.

\end{abstract}

\section{Introduction}\label{sec:introduction}
\begin{figure}[t]
    \centering
    \includegraphics[width=\columnwidth, trim=45 0 75 -10, clip]{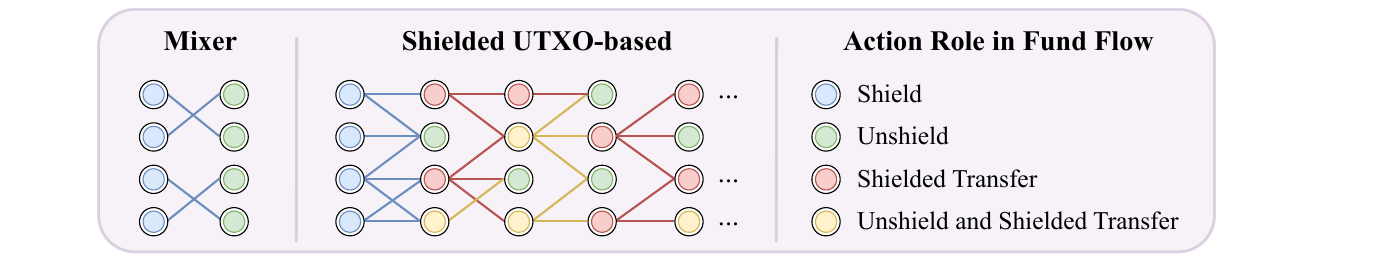}
    \caption{One-hop mixer matching versus recursive provenance in shielded \acs{UTXO}-based protocols.}
    \label{fig:introduction_comparison}
\end{figure}

Since the \ac{OFAC} sanctions on Tornado Cash in August 2022~\cite{OFAC2022TornadoCash}, on-chain demand for privacy has not disappeared; instead, protocol forms have continued to evolve~\cite{Baldimtsi2024SoK}. Privacy usage is shifting from coin mixers supporting separate deposit and withdrawal pools to shielded \ac{UTXO}-based protocols that support richer hidden-state operations. Railgun, a leading representative of the latter category, illustrates this trend: a recent Cambridge analysis reports that daily transactions on Tornado Cash Classic and Tornado Cash Nova fell by 97\% and 91\%, respectively, after the Tornado Cash sanctions, while Railgun's share of transactions among major Ethereum privacy protocols rose from 13\% in 2022 to 71\% in 2025 \cite{WuBear2026CambridgeMixers}. As of our observation cutoff, its \ac{TVL} was approximately \$80.23 million.\footnote{\acs{TVL} on July~30, 2026, from \href{https://defillama.com/protocol/railgun}{DefiLlama}.} Unlike mixers, these shielded \acs{UTXO}-based protocols let users move assets into the shielded pool (\emph{shield}), continue transferring, splitting, merging, and re-spending them inside the hidden state (\emph{shielded transfer}), and execute public release (\emph{unshield}) only when they need to return to the public state, for example, to withdraw or interact with public \ac{DeFi} protocols.
As \autoref{fig:introduction_comparison} shows, this protocol form changes the object of anonymity analysis. In mixer-style systems, fund flow is organized around the correspondence between deposit and withdrawal \cite{Wang2023Mixers,Meiklejohn2018Mobius,Heilman2017TumbleBit}, whereas in shielded \acs{UTXO}-based protocols, hidden state can be created, transferred, and re-spent, while a single spend transaction may aggregate multiple proofs. Accordingly, the feasible provenance sets underlying one \emph{unshield}, and hence its Anonymity Set Size, are no longer a one-hop pool-matching problem. Instead, provenance propagates across the note/\acs{UTXO}, proof, and transaction layers: the note layer provides candidate hidden-state objects, the proof layer captures private-state relations, and the transaction layer exposes entry and exit addresses.
Previous work has used address clustering, timing, value patterns, and other behavioral or attribution heuristics to reassess deployed privacy, including for Zcash, smart-contract mixers, Umbra, and, most recently, Railgun on Ethereum~\cite{Kappos2018Zcash,Wang2023Mixers,Kovacs2024Umbra,Huseynov2026Railgun}.
A distinct line derives eliminations from public ring-membership structure in Monero and related systems, sometimes combined with temporal or decoy-selection heuristics~\cite{Moser2018MoneroTrace,Vijayakumaran2023DM,Chow2023SustainableRings}.
To the best of our knowledge, this work provides the first systematic, non-heuristic analysis of how recursive provenance shapes transaction-level anonymity in deployed shielded \acs{UTXO}-based protocols for \acs{DeFi}.
Using only publicly observable protocol traces and constraints, we set the analysis boundary at the transaction level: the unit is a transaction that executes unshield, and its candidate sources are protocol-visible shielding addresses. This design isolates anonymity loss caused by the protocol's own public constraints. We neither cluster addresses into users or entities nor weight candidates probabilistically. The former typically requires heuristic or off-chain attribution and the latter an explicit probabilistic model. Separating these choices from protocol-visible constraints yields a non-heuristic baseline for either extension. Within this scope, we ask: \emph{from which shielding addresses could the funds released by a transaction that executes unshield have originated, and by how much can that address set be reduced?}

To answer this question, we design an empirical analysis that operates on executed traces and propagates constraints across layers. We organize the problem into three connected parts: first, the system model and privacy metrics; second, the proof-level shrinking of feasible hidden sources; third, the connection from these proof-level results to the final Anonymity Set Size at the transaction level. This organization reflects the actual distribution of public constraints in the target systems: hidden-state objects live at the note layer, private-state relations at the proof layer, and the visible anonymity outcome at the transaction layer.

\point{Layered system model and privacy metrics}
We first formalize a layered system model for shielded \acs{UTXO}-based protocols and define transaction-level Anonymity Set Size together with the proof-level Commitment Set Size used to compute it. Commitment Set Size captures how many historical commitments remain feasible as hidden inputs for a given proof. Anonymity Set Size captures how many shielding addresses could have originated the funds released by one transaction that executes \emph{unshield}. The model connects local hidden-source feasibility to final public release across the note, proof, and transaction layers. The formal model and metrics are developed in \autoref{sec:system}.

\point{Commitment-source pruning}
We then construct the Commitment Set of each proof and perform cumulative pruning over that set. Temporal order, tree number, historical roots, token compatibility, and value feasibility jointly determine whether one historical commitment can still serve as a feasible hidden input to the current proof. Starting from the temporal baseline, we apply these constraints and propagate their consequences to shrink each proof's feasible hidden sources. No address clustering, behavioral patterns, or external attribution signals are used. The full pipeline is presented in \autoref{sec:deanonymization}.

\point{Closure-based lifting to transaction anonymity}
Finally, we start from the Commitment Set obtained for each proof after cumulative pruning and propagation and trace its feasible provenance backward along the hidden-state history. These commitments may be introduced directly by \emph{shield} or created by later \emph{shielded transfer} and \emph{unshield and shielded transfer} steps, while the final Anonymity Set Size identifies the shielding addresses through which the released funds could originally have entered the pool. We therefore connect the still-reachable commitments to historical shield transactions and their shielding addresses, and aggregate these address sets within each transaction. We instantiate this process on the complete histories of all four Railgun production deployments and the five most active independent pools in Hinkal's largest coordinated \ac{EVM}-compatible deployment generation by aggregate commitments. Across these analyses, we cover 313{,}492 proofs and 259{,}049 spend transactions, of which 186{,}356 are unshielding spend transactions. Relative to their respective temporal baselines, the mean Anonymity Set Size shrinks by 40.1\%--59.0\% across the Railgun chains and by 48.9\%--57.4\% across the Hinkal chains. The final set contains at most ten addresses in 2{,}968 Railgun and 711 Hinkal transactions, including 920 and 308 singleton cases, respectively. From these measurements, we distill interpretable anonymity-loss patterns, explain when very small sets emerge, and derive practical implications for user behavior, protocol interfaces, and privacy-protocol design, as developed in \autoref{sec:empirical} and \autoref{sec:discussion}.

\point{Contributions}
This work makes the following contributions:
\begin{itemize}
    \item We formalize a layered system model for shielded \acs{UTXO}-based protocols and define transaction-level Anonymity Set Size together with the proof-level Commitment Set Size used to compute it.
    \item We introduce a non-heuristic recursive provenance analysis combining four-stage pruning with cross-layer lifting to trace feasible hidden-state histories to shielding addresses and derive transaction-level Anonymity Set Size.
    \item We analyze the complete on-chain histories of all four Railgun production deployments and five independent Hinkal pools, finding mean Anonymity Set Size reductions of 40.1\%--59.0\% and 48.9\%--57.4\%, respectively, relative to their temporal baselines.
    \item We identify an eight-pattern anonymity-loss taxonomy across proof-level specialization, proof-structure effects, and transaction-level aggregation, support it with case studies, and distill it into practical implications for user behavior, privacy-protocol interfaces and design.
\end{itemize}

\section{Background}\label{sec:background}

Permissionless blockchains such as Ethereum execute transactions in a globally agreed order on a shared append-only ledger, yielding a publicly inspectable trace \cite{Wood2014Ethereum,Beres2021BlockchainWatching}. Smart contracts store protocol state and enforce state transitions, but their calls, execution order, state updates, and events remain visible even when cryptography hides the underlying asset state \cite{Baldimtsi2024SoK}. \acs{DeFi}-facing privacy protocols must preserve hidden state in this public setting. We study \emph{shielded \acs{UTXO}-based protocols for \acs{DeFi}}, a target class defined by five features that distinguish it from other blockchain privacy protocols, such as mixers and stealth address protocols.

\point{Commitment-Nullifier State}
At the state-model level, the target class represents value as discrete notes rather than mutable account balances \cite{Baldimtsi2024SoK}. Creating a note posts a hiding commitment to an append-only accumulator, typically a Merkle tree \cite{Merkle1988DigitalSignature}. To spend privately, a user publishes a note-derived nullifier and uses a \ac{ZKP} to prove knowledge of and authorization for a note whose commitment is covered by a recorded public Merkle root, as well as the nullifier's correctness, without revealing the consumed commitment \cite{BenSasson2014Zerocash,Rondelet2019ZETH}. The published nullifier prevents double spending. Deployed systems typically instantiate this proof as a \ac{SNARK} \cite{BenSasson2014Zerocash,Groth2016SNARK,Guo2024ZKBenchmark}. This commitment-nullifier discipline defines the hidden-state machine later abstracted in our model. Umbra instead derives receiver-specific stealth addresses without such a commitment-nullifier pool \cite{UmbraDocs2026}.

\point{Shielded Transfer}
A private transition can consume existing notes and create new ones, while a \ac{ZKP} proves input, output, and protocol consistency without revealing the witnesses \cite{BenSasson2014Zerocash,Rondelet2019ZETH}. Unlike withdrawal-oriented mixers (see \autoref{fig:introduction_comparison}), value can remain hidden, recur as new notes, and continue through later private transitions \cite{RailgunDocs2026,TornadoCashNovaDocs2026,zkBobDocs2026,ZcashProtocolSpecification2026}.

\point{Token-Encoded Notes}
Supporting multiple assets requires locating asset identity. A single-asset pool fixes it by deployment; the target class instead encodes it in each note, allowing one hidden-state machine to represent multiple assets and enforcing token consistency in private transitions \cite{RailgunDocs2026,HinkalWhitepaper2023,NocturneDocs2026}. Note encoding does not determine public visibility: an implementation may expose the token directly or reveal only transaction-specific constraints from which it can sometimes be inferred. Such token information can induce publicly observable pruning constraints, whereas single-asset pools factor this distinction out at deployment time \cite{TornadoCashClassicDocs2026,zkBobDocs2026}.

\point{Recipient Discovery}
To recover and spend a note created by another user, the recipient must identify its public commitment. The sender typically publishes recipient-bound data derived from the recipient's public key with the commitment, allowing the recipient to scan with a private key \cite{RailgunDocs2026,HinkalWhitepaper2023}; otherwise, recovery data must be conveyed off-protocol \cite{TornadoCashClassicDocs2026}.

\point{\acs{DeFi}-Facing Design}
We call a design \acs{DeFi}-facing when the protocol itself lets shielded state feed broader on-chain financial activity, rather than only allowing withdrawal followed by external use. Railgun unshields assets into the Relay Adapt contract, executes external calls, and shields the result back in one transaction \cite{RailgunDocs2026}. Hinkal's pool similarly executes external calls using shielded state, supporting private swaps, lending, and staking; transactions may be submitted through relayers \cite{HinkalWhitepaper2023}. Tornado Cash Nova instead centers on deposit, in-pool transfer, and withdrawal \cite{TornadoCashNovaDocs2026}.

\autoref{tab:protocol_scope} compares representative protocols along these dimensions and motivates three scopes:
\begin{itemize}
    \item \textbf{Shielded \acs{UTXO}-based Protocols for \acs{DeFi}} combine all five features and form our direct modeling target; this class includes Railgun, Hinkal, and Nocturne.\footnote{Nocturne \href{https://x.com/nocturne_xyz/status/1798420775739023466}{ceased operations in 2024}.}
    \item \textbf{Adjacent Commitment-Nullifier Designs}, including Tornado Cash Nova, zkBob, and Zcash, share much of the hidden-state machinery but differ in token encoding or \acs{DeFi}-facing scope. Tornado Cash Classic is a simpler mixer within the broader commitment-nullifier family.
    \item \textbf{Other Blockchain Privacy Protocols}, such as the stealth address protocol Umbra, lie outside this family and require a distinct anonymity model because their state transitions expose different public constraints.
\end{itemize}

\begin{table}[t]
    \centering
    \tiny
    \setlength{\tabcolsep}{2.0pt}
    \renewcommand{\arraystretch}{1.08}
    \newcommand{\scopeyes}{\raisebox{0.08ex}{\scalebox{1.15}{$\bullet$}}}
    \newcommand{\scopeno}{\raisebox{0.08ex}{\scalebox{1.15}{$\circ$}}}
    \caption{Mechanism-level feature comparison of representative blockchain privacy protocols ($\bullet$ = yes, $\circ$ = no).}
    \label{tab:protocol_scope}
    \resizebox{\columnwidth}{!}{%
        \begin{tabular}{lcccccccc}
            \toprule
            \multirow{2}{*}{Feature}
                                       & \multirow{2}{*}{\shortstack{Railgun                                                                                                \\\cite{RailgunDocs2026}}}
                                       & \multirow{2}{*}{\shortstack{Hinkal                                                                                                 \\\cite{HinkalWhitepaper2023}}}
                                       & \multirow{2}{*}{\shortstack{Nocturne                                                                                               \\\cite{NocturneDocs2026}}}
                                       & \multicolumn{2}{c}{Tornado Cash}
                                       & \multirow{2}{*}{\shortstack{zkBob                                                                                                  \\\cite{zkBobDocs2026}}}
                                       & \multirow{2}{*}{\shortstack{Zcash                                                                                                  \\\cite{ZcashProtocolSpecification2026}}}
                                       & \multirow{2}{*}{\shortstack{Umbra                                                                                                  \\\cite{UmbraDocs2026}}}                                                                                               \\
            \cmidrule(lr){5-6}
                                       &                                      &           &           & \shortstack{Classic\\\cite{TornadoCashClassicDocs2026}}   & \shortstack{Nova\\\cite{TornadoCashNovaDocs2026}}      &           &           &           \\
            \midrule
            Commitment-Nullifier State & \scopeyes                            & \scopeyes & \scopeyes & \scopeyes           & \scopeyes & \scopeyes & \scopeyes & \scopeno  \\
            Shielded Transfer          & \scopeyes                            & \scopeyes & \scopeyes & \scopeno            & \scopeyes & \scopeyes & \scopeyes & \scopeno  \\
            Token-Encoded Notes        & \scopeyes                            & \scopeyes & \scopeyes & \scopeno            & \scopeno  & \scopeno  & \scopeno  & \scopeno  \\
            Recipient Discovery        & \scopeyes                            & \scopeyes & \scopeyes & \scopeno            & \scopeyes & \scopeyes & \scopeyes & \scopeyes \\
            \acs{DeFi}-Facing Design   & \scopeyes                            & \scopeyes & \scopeyes & \scopeno            & \scopeno  & \scopeyes & \scopeno  & \scopeno  \\
            \bottomrule
        \end{tabular}%
    }
\end{table}

\section{System Model and Privacy Metrics}\label{sec:system}

For the target class of \autoref{sec:background}, this section formalizes the protocol lifecycle across note, proof, and transaction layers and defines Commitment Set Size and Anonymity Set Size with their temporal baselines. The transaction layer is our public observation boundary, whose visible units are on-chain entry and exit addresses; the metrics therefore characterize address-level ambiguity over shielding addresses. Extending the address metric to users or entities requires address attribution, often using off-chain information or heuristics, and lies outside our protocol-visible analysis. Because one user may control multiple shielding addresses, address-level set sizes generally upper-bound user-level sizes, except when multiple users share one shielding address.

\subsection{System Model}\label{ssec:system_model}

\point{Shielded Pool}
A shielded pool $\mathcal{P}$ is the protocol-maintained private state over supported assets and exposes four actions.

\point{\fcirc{1}~Shield}
One asset type, such as an ERC-20 token or \ac{NFT}, is moved from the public state into $\mathcal{P}$. The protocol creates notes that encode the corresponding spending rights in the pool.

\point{\fcirc{2}--\fcirc{4} Spending}
The user spends notes of a single asset type inside $\mathcal{P}$. Recording their nullifiers consumes them, and the spent value is processed as follows.

\begin{itemize}
      \item \fcirc{2}~\textbf{Unshield:} All spent value is unshielded from $\mathcal{P}$ to an external address, moving assets back to the public state.
      \item \fcirc{3}~\textbf{Shielded Transfer:} All spent value remains within $\mathcal{P}$ and is reassigned within the shielded state.
      \item \fcirc{4}~\textbf{Unshield and Shielded Transfer:} The spent value is split such that one part is unshielded to an arbitrary external address, while the remaining part stays within the shielded state of $\mathcal{P}$.
\end{itemize}

\subsubsection{Note (\texorpdfstring{\acs{UTXO}}{UTXO}) Layer}\label{ssec:note}

The note (\acs{UTXO}) layer contains the basic state objects of the shielded pool and their permanent on-chain execution trace.

\point{Note}
A note $N$ represents the abstract state object created and consumed by the four actions. Each note is created exactly once and can be consumed at most once. Notes are not stored on-chain and are represented by commitments.

\point{Commitment}
Each note $N$ maps one-to-one to an on-chain commitment $c$. A commitment encodes the note's fixed asset attributes. A commitment remains on-chain regardless of whether the underlying note has already been consumed.

\point{Nullifier}
When a note represented by commitment $c$ is consumed, the protocol records a unique public nullifier deterministically derived from that note's secrets. The nullifier prevents double spending without revealing $c$ or note attributes.

\subsubsection{Proof Layer}\label{ssec:proof}

The proof layer captures the same-token input--output relations through which $\mathcal{P}$ processes spent notes.

\point{Proof}
A proof, denoted by $\pi_i$, models one such relation; a cryptographic proof may jointly authenticate multiple such relations. It consumes existing notes, records their nullifiers, and may create new commitments. All notes consumed by $\pi_i$ share one token type, and their values must satisfy the protocol's balance relation. Each proof may include a plaintext output $O_i=(\tau_i,U_i)$, whose token type and value describe assets released from $\mathcal{P}$ to the public state; $O_i=\varnothing$ denotes a purely internal proof. When present, its revealed semantics must match the token type and released value induced by the consumed notes. Let $m_i$ and $k_i$ denote the numbers of notes consumed and commitments newly created by $\pi_i$, respectively. We distinguish three proof types corresponding to Actions~\fcirc{2}--\fcirc{4}: Proof~\fcirc{2}, Proof~\fcirc{3}, and Proof~\fcirc{4}.
\begin{itemize}
      \item \textbf{Unshield proof (Proof~\fcirc{2}).}
            The proof consumes existing notes and releases all spent value through a plaintext output $O_i=(\tau_i,U_i)$.
            No new commitments are created.

      \item \textbf{Shielded-transfer proof (Proof~\fcirc{3}).}
            The proof consumes existing notes, releases no value from $\mathcal{P}$, and keeps all spent value within the shielded state, with $O_i=\varnothing$ throughout.

      \item \textbf{Unshield-and-shielded-transfer proof (Proof~\fcirc{4}).}
            The proof consumes existing notes, releases part of the spent value through a plaintext output $O_i=(\tau_i,U_i)$, and keeps the rest within the shielded state.
\end{itemize}

Commitments in $\mathcal{P}$ are \emph{shield commitments} introduced by Action~\fcirc{1} or \emph{proof-created commitments} introduced by Proof~\fcirc{3} or Proof~\fcirc{4}. Each proof-created commitment has a unique creating proof, denoted $\pi(c)$. This classification identifies how a commitment is created rather than requiring mutually exclusive funding provenance: when a transaction combines shielding and spending, a proof-created commitment may depend on a public entry and earlier shielded notes.

\subsubsection{Transaction Layer}\label{ssec:transaction}

The transaction layer defines the execution boundary and temporal order of $\mathcal{P}$ state transitions. Commitments created within a transaction cannot serve as inputs to its proofs; only those present before execution are eligible. Shield and spend roles in the public trace need not be mutually exclusive.

\point{Shield transaction}
A \emph{shield transaction} corresponds to Action~\fcirc{1}: it inserts shield commitments into $\mathcal{P}$.

\point{Shielding address}
The shielding address of a shield transaction is the public address from which assets enter $\mathcal{P}$---the entry point to hidden state.

\point{Spend transaction}
A \emph{spend transaction} aggregates one or more proofs and is either:
\begin{itemize}
      \item \textbf{Internal spend transaction:}
            every constituent proof is Proof~\fcirc{3}, so no assets are released to the public state.
      \item \textbf{Unshielding spend transaction:}
            at least one constituent proof is Proof~\fcirc{2} or Proof~\fcirc{4}. Only this class carries the Anonymity Set Size defined below.
\end{itemize}

\point{Unshielding address}
The public address to which a Proof~\fcirc{2} or Proof~\fcirc{4} releases assets from $\mathcal{P}$ is its unshielding address, marking the exit point from hidden state.

\begin{figure}[t]
    \centering
    \includegraphics[
        width=0.7\columnwidth,
        trim=55 90 150 10,
        clip
    ]{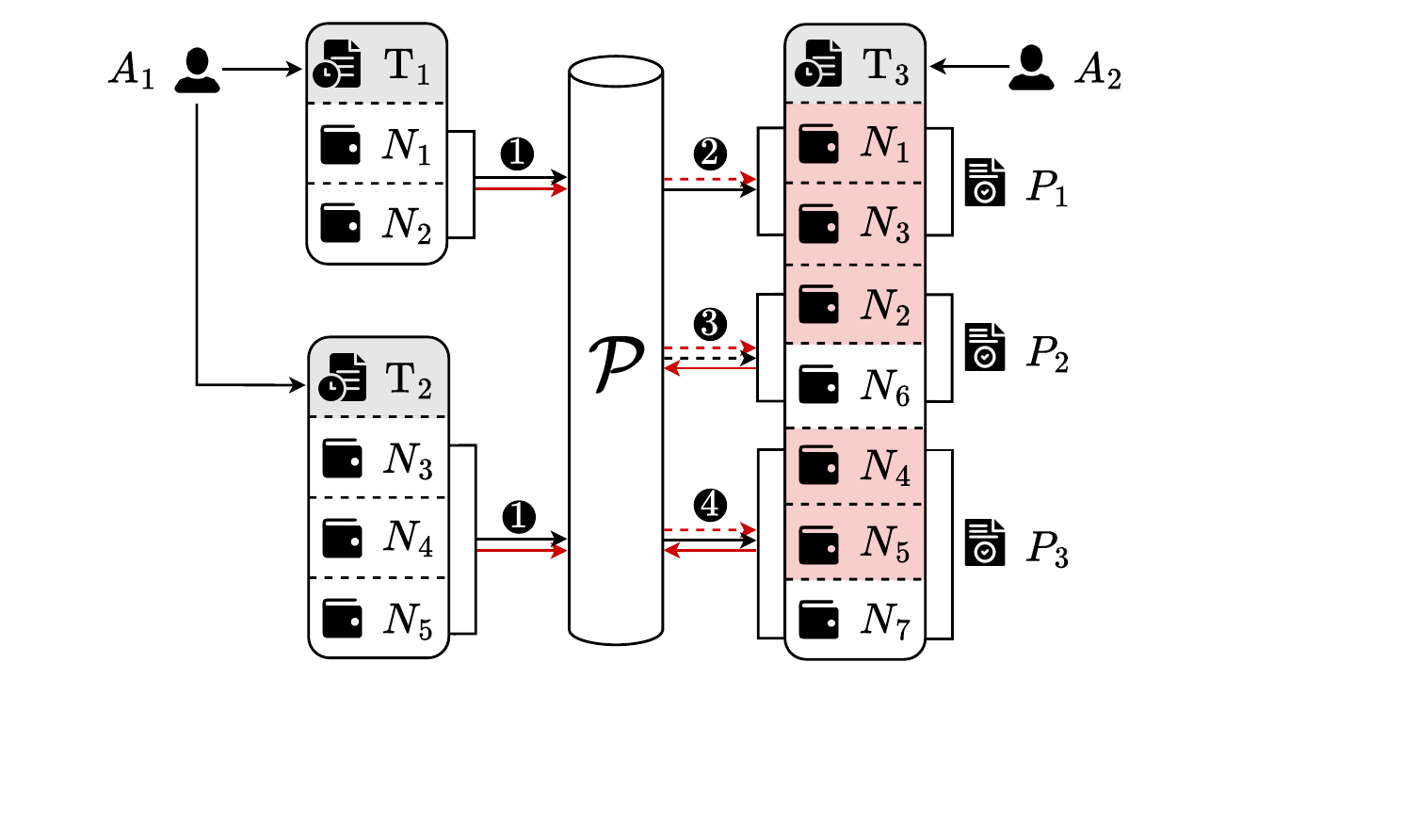}
    \caption{Illustrative transaction-intent view of the system model for shielded pool $\mathcal{P}$. Transactions $T_1$ and $T_2$ are initiated by address $A_1$, and $T_3$ by address $A_2$. The symbol $\rightarrow$ denotes fund transfers, while $\dashrightarrow$ denotes ownership transfers within $\mathcal{P}$. The red arrow $\textcolor{red}{\rightarrow}$ indicates note insertion into $\mathcal{P}$, and the red dashed arrow $\textcolor{red}{\dashrightarrow}$ indicates note nullification. Action~\fcirc{1}: Shield; \fcirc{2}: Unshield; \fcirc{3}: Shielded Transfer; \fcirc{4}: Unshield and Shielded Transfer. Red-background notes are consumed by the corresponding proof; unshaded notes are newly created.}
    \label{fig:system_model}
\end{figure}

\autoref{fig:system_model} instantiates the three-layer model. Shield transactions $T_1$ and $T_2$ from $A_1$ create $N_1,N_2$ and $N_3,N_4,N_5$, respectively. Transaction $T_3$, initiated by $A_2$, aggregates $P_1,P_2,P_3$, which are Proof~\fcirc{2}, Proof~\fcirc{3}, and Proof~\fcirc{4}, respectively. $P_1$ consumes $N_1,N_3$ and creates none; $P_2$ consumes $N_2$ and creates $N_6$; $P_3$ consumes $N_4,N_5$, creates $N_7$, and releases value to the public state. The example connects note creation and consumption to proof aggregation and shows why outputs created within $T_3$ cannot feed another proof in $T_3$.

\subsection{Privacy Metrics}\label{ssec:privacy_metrics}

The metrics below count the commitments or shielding addresses that remain feasible under protocol-visible constraints, without ranking them by likelihood. A probability-weighted view requires an explicit posterior based, for example, on behavioral priors, attribution models, or off-chain information; these sets can then serve as candidate spaces for entropy-based anonymity metrics~\cite{Diaz2003MeasuringAnonymity,Serjantov2003InfoMetric}.

\point{Commitment Set Size $|\Omega(i)|$}
For a proof $\pi_i$, $\Omega(i)$ is the set of commitments feasible as its hidden inputs; $|\Omega(i)|$ is the Commitment Set Size. The metric has proof granularity because all nullifiers recorded by the same proof share the same constraints and hence the same Commitment Set.

\point{Temporal baseline $\Omega_0(i)$}
For a proof $\pi_i$, the temporal baseline $\Omega_0(i)$ contains all commitments present when its containing spend transaction begins execution. In \autoref{fig:system_model}, the proof in $T_3$ that consumes $N_4$ and $N_5$ has the five historical commitments as its baseline; $N_6$ and $N_7$, created within $T_3$, are excluded, so both nullifiers share that baseline.

\point{Anonymity Set Size $|\mathcal{A}(T)|$}
For an unshielding spend transaction $T$, $\mathcal{A}(T)$ is the set of shielding addresses from which $T$'s released funds could have originated:
\begin{equation}\label{eq:anonymity_set}
      \mathcal{A}(T)=\bigcup_{\substack{\pi_i\in T\\O_i\neq\varnothing}}\mathsf{Addr}\!\big(\Omega(i)\big),
\end{equation}
where $\mathsf{Addr}(\Omega(i))$ denotes the potential shielding addresses obtained by tracing the commitments in $\Omega(i)$, together with any protocol-visible predecessor relation attached directly to $\pi_i$, back to the shield transactions from which they could originate. Its size $|\mathcal{A}(T)|$ is the Anonymity Set Size.

\point{Temporal baseline $\mathcal{A}_0(T)$}
For an unshielding spend transaction $T$, the temporal baseline $\mathcal{A}_0(T)$ contains the shielding addresses of all preceding shield transactions. In \autoref{fig:system_model}, $T_1$ and $T_2$ both use $A_1$, so $\mathcal{A}_0(T_3)=\{A_1\}$.
We report the relative reduction $\mathrm{red}_{\mathcal{A}}(T)=1-|\mathcal{A}(T)|/|\mathcal{A}_0(T)|$.

\section{Non-Heuristic Provenance Analysis}\label{sec:deanonymization}

This section instantiates the metrics defined in \autoref{ssec:privacy_metrics}. Cumulative pruning constructs $\Omega(i)$ from public traces, and lifting maps the survivors to shielding addresses and unions them within each unshielding spend transaction to obtain $\mathcal{A}(T)$.

\subsection{Commitment Set Construction}\label{ssec:note_source_pruning}

\begin{figure*}[t]
    \centering
    \includegraphics[width=0.98\textwidth]{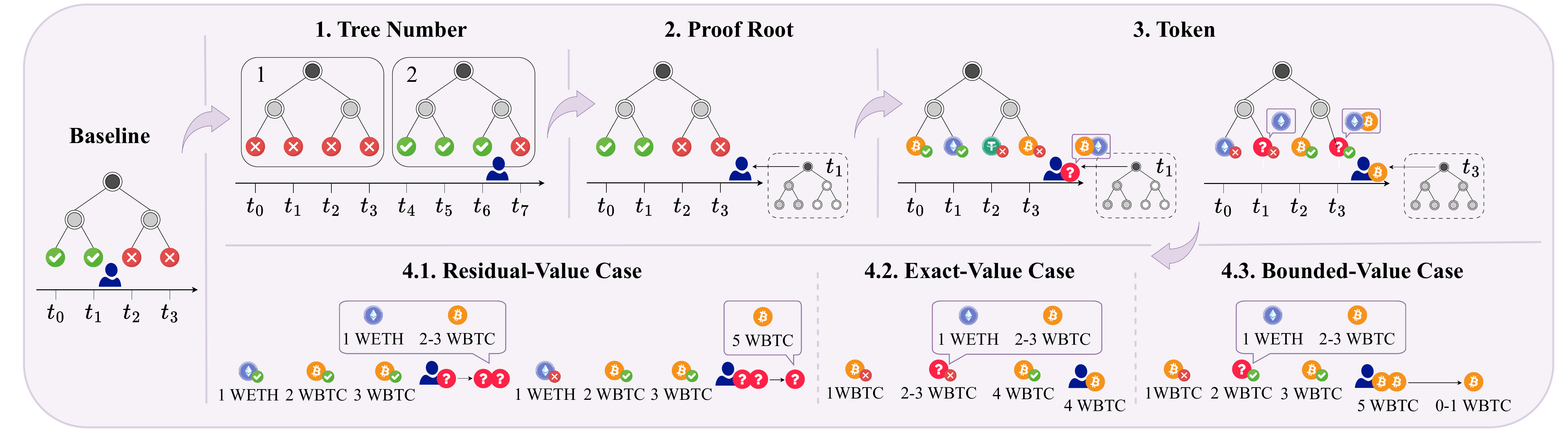}
    \caption{Leaf insertions $t_0$--$t_7$ and dashed Merkle-tree snapshots at $t_1,t_3$ illustrate commitment-set pruning. Candidate leaves carry applicable token/value labels; the blue user marks the spend; green ticks/red crosses denote retained/pruned candidates.}
    \label{fig:strategy_overview}
\end{figure*}

Starting from the temporal baseline $\Omega_0(i)$ defined in \autoref{ssec:privacy_metrics}, the pipeline evaluates four cumulative stages: tree number, proof-root coverage, token consistency, and value feasibility, denoted by $S_\mathrm{treeNumber}$, $S_\mathrm{proofRoot}$, $S_\mathrm{token}$, and $S_\mathrm{value}$. They yield $\Omega_1(i)$ through $\Omega_4(i)$ before cross-proof propagation; a structurally inapplicable stage acts as the identity. \autoref{fig:strategy_overview} illustrates each constraint on $\pi_i$ separately. These snapshots report one feasible-set construction, not extra pruning assumptions: $\Omega_4(i)$ contains baseline candidates not ruled out by applicable public constraints. The tree/root constraints are intersective; tree-first evaluation is efficient because root coverage is tree-local. The token/value tail is representation-dependent: $S_\mathrm{token}$ constructs $\Theta_i$ before $S_\mathrm{value}$ tests each branch for joint token--value feasibility. Thus, order affects snapshots and attribution, not the final local feasible set used by propagation.

\subsubsection{Strategy 1: Tree Number}\label{ssec:s1}

Panel~1 of \autoref{fig:strategy_overview} shows the simplest structural cut. A private spend proves membership in one public Merkle tree, so time-eligible commitments in other trees are impossible; in the example, the reference to tree~2 excludes the earlier leaves in tree~1. \autoref{fig:alg_s1_tree} formalizes this public comparison, retaining commitments in the same tree as $\pi_i$ to form $\Omega_1(i)$ without token or value reasoning.

\subsubsection{Strategy 2: Proof Root}\label{ssec:s2}

Panel~2 of \autoref{fig:strategy_overview} refines this boundary within the selected tree. A spend is verified against one public root snapshot, so any commitment inserted after its largest covered index is too new. In the example, the leaf added between $r_0$ and $r_1$ becomes eligible only when the spend references $r_1$. \autoref{fig:alg_s2_root} formalizes the prefix test $\ell(c)\leq\ell_{\max}(\pi_i)$, whose survivors form $\Omega_2(i)$.

\subsubsection{Strategy 3: Token}\label{ssec:s3}

Panel~3 of \autoref{fig:strategy_overview} shows the first semantic cut: which token interpretations can explain the proof's hidden inputs. In the left example, the token is not revealed, so BTC and ETH inherited from the remaining candidates stay feasible, while USDT does not. In the right example, under the BTC branch, an ETH-only leaf drops out, whereas a commitment still compatible with BTC and ETH survives.

\begin{figure}[t]
    \centering
    \begin{minipage}[c]{0.485\columnwidth}
        \input{algorithm/alg_s1_tree}

        \vspace{0.8\baselineskip}

        \input{algorithm/alg_s2_proofroot}
    \end{minipage}\hfill
    \begin{minipage}[c]{0.485\columnwidth}
        \input{algorithm/alg_s3_token}
    \end{minipage}
\end{figure}

\autoref{fig:alg_s3_token} formalizes this test. For each candidate $c$, $\Lambda(c)$ is the token set permitted by its public creation trace and, if $c$ is proof-created, the current feasible output state of its creating proof $\pi(c)$; propagation can only narrow it. Intersecting these labels with the token set $\mathcal{T}_i$ permitted by $\pi_i$, the algorithm removes branches with fewer than the required $m_i$ inputs and returns the survivors $\Theta_i$ and their commitments $\Omega_3(i)$.

\subsubsection{Strategy 4: Value}\label{ssec:s4}

Panels~4.1--4.3 of \autoref{fig:strategy_overview} show where arithmetic starts pruning. With neither a public token nor a public output value, each token branch must independently support the required inputs; one leaf cannot explain a two-input spend. With an exact public amount, only candidates whose values can participate in attaining it survive. With 5~WBTC revealed and one hidden output, the total input may instead lie within a bounded interval, excluding leaves entirely outside it.

\autoref{fig:alg_s4_value} evaluates value feasibility within each branch $t\in\Theta_i$ retained by $S_\mathrm{token}$. The filter $\Phi_{i,t}$ applies the public balance relation of $\pi_i$ to produce the feasible branch slice $\Omega_4^t(i)$. For Panels~4.1--4.3, the targets are $[0,k_iV_{\max}]$, $[U_i,U_i]$, and $[U_i,U_i+k_iV_{\max}]$, where $k_i$ is the number of commitments created by $\pi_i$, $U_i$ is its revealed output value when present, and $V_{\max}$ bounds the value of one created commitment. The extractor $\Gamma_{i,t}$ then returns the branch-local mandatory set $M^t(i)$, whose leaves occur in every remaining feasible explanation. The surviving branches form $\Theta_i^\star$; the union of their slices gives $\Omega_4(i)$, while the intersection of their mandatory sets gives $M(i)$.

This branch-local operator is conservative: it removes a candidate only when the public token and value constraints rule out its participation; an unresolved value is insufficient. It does not enumerate every concrete $m_i$-input tuple during each fixpoint update, so a candidate may remain even if exhaustive joint solving could eliminate it. The reported sets therefore upper-bound the exact jointly feasible candidate sets.

\subsubsection{Cross-Proof Global Propagation}\label{ssec:ss}

The four local strategies treat each proof in isolation, but proofs can constrain one another in two ways: a mandatory input is consumed and unavailable to other compatible proof domains, while a created commitment may carry a narrower token or value interpretation into later private transitions. Cross-proof propagation follows the public state-transition semantics and revisits only affected proofs until no state can shrink further. Operationally, propagation keeps the structural prefix from Strategies~1 and~2 fixed and recomputes only the token--value state after a newly mandatory input or a narrowed label on a created commitment. Such contractions reach a fixpoint whose final $\Omega(i)$ and token-guarded family $\Sigma(i)$ feed closure lifting; \appref{app:combined_propagation} gives the formal procedure.

\subsection{Closure-Based Lifting to Anonymity Sets}\label{ssec:closure_lifting}

Local pruning identifies commitments that can explain each analyzed spend, but not their entry origins. Closure-based lifting recursively follows publicly determined shielded-state dependencies from each survivor and any predecessor exposed directly by the analyzed proof relation, yielding the historical shield transactions from which the released funds could have originated.

The backward operator $\Pi_\mathrm{closure}$ traces the final Commitment Set $\Omega(i)$, under its surviving token-guarded family $\Sigma(i)$, to its feasible terminal shield transactions $\mathcal{E}(i)$:
\begin{equation}\label{eq:closure_origins}
    \mathcal{E}(i)=\Pi_\mathrm{closure}(\pi_i,\Omega(i),\Sigma(i)).
\end{equation}
Mapping the transactions in $\mathcal{E}(i)$ to their shielding addresses yields $\mathsf{Addr}(\Omega(i))$ in \autoref{eq:anonymity_set}. To evaluate \autoref{eq:closure_origins}, \autoref{fig:alg_closure_lifting} keeps one backward queue per analyzed proof; a state $(j,t)$ traces $\pi_j$ under token guard $t$. The sets $\mathsf{Entry}_{\pi}(j)$ and $\mathsf{Pred}_{\pi}(j)$ contain terminal shield transactions and guarded predecessor states exposed directly by $\pi_j$; both are empty if its trace exposes no such edge. A direct predecessor remains traversable without a commitment input.

\begin{figure}[t]
    \centering
    \input{algorithm/alg_s4_value}\hfill
    \input{algorithm/alg_closure_lifting}
\end{figure}

For each candidate $c$, $\mathsf{Entry}(c,t)$ and $\mathsf{Pred}(c,t)$ respectively return the terminal shield transactions and earlier guarded proof states permitted by its public creation trace. Both retain all publicly indistinguishable alternatives and may be non-empty together. A proof-created commitment follows $\pi(c)$ under the same guard unless the public trace determines another token relation. Each guarded state is expanded once; the final loop then computes $\mathcal{A}(T)$ via \autoref{eq:anonymity_set}.

\section{Results}\label{sec:empirical}

This section evaluates the pipeline of \autoref{sec:deanonymization} using the inclusive cutoff July~30,~2026 at 23:59:59 UTC. For Railgun, we reconstruct and analyze the complete on-chain histories of all four production deployments---Ethereum, Arbitrum, Polygon, and BNB Chain---from the initial deployment on each chain through the cutoff, covering every commitment, nullifier, proof, shield transaction, spend transaction, and Merkle-tree rollover recorded within the window. Hinkal, by contrast, has a highly fragmented deployment landscape: among 39 independently deployed pool states across ten chains, 37 are nonempty and 23 of those contain fewer than 1{,}000 commitments. Because each deployment maintains an independent commitment tree and pool state, these pools cannot be concatenated into one continuous anonymity domain. \appref{app:hinkal_deployments} inventories this landscape. We therefore select the coordinated \acs{EVM}-compatible generation deployed on June~27,~2025, which has the largest aggregate commitment population among Hinkal's coordinated cross-chain generations in the window, and evaluate its five most active pools on Ethereum, Arbitrum, Polygon, Base, and Optimism.

To keep the cross-protocol comparison consistent with \autoref{ssec:proof}, we count proofs at the granularity of same-token input--output relations. A Railgun JoinSplit proof contains one such relation, whereas a single Hinkal cryptographic proof may jointly authenticate several; accordingly, each input-bearing same-token relation within a Hinkal cryptographic proof is counted separately. \appref{app:protocol-instantiations} gives the implementation mapping and evidence boundary, and \appref{app:analysis_validation} documents the full-trace reconstruction and fail-closed analysis checks. \autoref{tab:dataset_overview} and \autoref{tab:hinkal_dataset_overview} summarize the resulting datasets. The Hinkal Anonymity Set Size evaluation covers contract-identified recipient withdrawals: outputs that are provably fee-only under the contract's execution semantics are excluded, while transactions containing both a recipient withdrawal and a fee remain included.

\begin{table}[t]
    \caption{Cross-Protocol Dataset Overview.}
    \label{tab:cross_protocol_dataset_overview}
    \centering
    \scriptsize
    \setlength{\tabcolsep}{1.70pt}
    \renewcommand{\arraystretch}{1.06}
    \begin{subtable}{\linewidth}
    \caption{Railgun.}
    \label{tab:dataset_overview}
    \centering
    \begin{adjustbox}{max width=\linewidth}
    \begin{tabular}{@{}lrrrrrrr@{}}
        \toprule
        Chain & \multicolumn{1}{c}{Commit.$^{\dagger}$} & \multicolumn{1}{c}{Nullif.} & \multicolumn{1}{c}{Proofs} & \multicolumn{1}{c}{Shield Txs} & \multicolumn{1}{c}{Shield Addrs.$^{\ddagger}$} & \multicolumn{1}{c}{Spend Txs} & \multicolumn{1}{c}{Unshield Txs} \\
        \midrule
        \multirow{4}{*}{Ethereum}
          & 65{,}536 & 61{,}804 & 33{,}124
          & \multirow{4}{*}{57{,}970} & \multirow{4}{*}{28{,}486}
          & \multirow{4}{*}{105{,}608} & \multirow{4}{*}{85{,}157} \\
          & 65{,}535 & 60{,}984 & 33{,}478 & & & & \\
          & 65{,}536 & 51{,}428 & 31{,}780 & & & & \\
          & 52{,}695 & 43{,}257 & 24{,}880 & & & & \\
        \cmidrule(lr){1-8}
        \multirow{2}{*}{Arbitrum}
          & 65{,}535 & 60{,}810 & 34{,}605
          & \multirow{2}{*}{30{,}049} & \multirow{2}{*}{10{,}911}
          & \multirow{2}{*}{50{,}933} & \multirow{2}{*}{42{,}256} \\
          & 50{,}782 & 38{,}436 & 26{,}196 & & & & \\
        \cmidrule(lr){1-8}
        \multirow{2}{*}{Polygon}
          & 65{,}536 & 58{,}122 & 34{,}244
          & \multirow{2}{*}{18{,}318} & \multirow{2}{*}{5{,}516}
          & \multirow{2}{*}{35{,}262} & \multirow{2}{*}{27{,}719} \\
          & 20{,}676 & 16{,}549 & 11{,}209 & & & & \\
        \cmidrule(lr){1-8}
        BNB Chain & 45{,}203 & 39{,}808 & 21{,}604
                  & 12{,}028 & 5{,}493 & 18{,}667 & 14{,}708 \\
        \midrule
        Total & 497{,}034 & 431{,}198 & 251{,}120
              & 118{,}365 & 47{,}576 & 210{,}470 & 169{,}840 \\
        \bottomrule
    \end{tabular}
    \end{adjustbox}
    \vspace{2pt}

    \begin{minipage}{\linewidth}
        \raggedright
        \footnotesize
        $^{\dagger}$ For each chain, successive rows represent trees $0,1,\ldots$ in order.
    \end{minipage}
    \end{subtable}

    \vspace{6pt}

    \begin{subtable}{\linewidth}
    \caption{Hinkal.}
    \label{tab:hinkal_dataset_overview}
    \centering
    \begin{adjustbox}{max width=\linewidth}
    \begin{tabular}{@{}lrrrrrrr@{}}
        \toprule
        Chain & \multicolumn{1}{c}{Commit.} & \multicolumn{1}{c}{Nullif.} & \multicolumn{1}{c}{Proofs} & \multicolumn{1}{c}{Shield Txs} & \multicolumn{1}{c}{Shield Addrs.$^{\ddagger}$} & \multicolumn{1}{c}{Spend Txs} & \multicolumn{1}{c}{Unshield Txs} \\
        \midrule
        Ethereum & 5{,}997  & 5{,}683  & 5{,}167  & 2{,}629  & 780   & 4{,}169  & 1{,}884 \\
        Arbitrum & 8{,}579  & 8{,}187  & 7{,}009  & 4{,}446  & 630   & 5{,}576  & 1{,}613 \\
        Polygon  & 13{,}692 & 12{,}863 & 10{,}302 & 6{,}154  & 655   & 8{,}276  & 3{,}156 \\
        Base     & 24{,}453 & 23{,}586 & 19{,}712 & 12{,}267 & 2{,}327 & 15{,}431 & 5{,}851 \\
        Optimism & 25{,}132 & 24{,}180 & 20{,}182 & 12{,}456 & 437   & 15{,}127 & 4{,}012 \\
        \midrule
        Total    & 77{,}853 & 74{,}499 & 62{,}372 & 37{,}952 & 4{,}123 & 48{,}579 & 16{,}516 \\
        \bottomrule
    \end{tabular}
    \end{adjustbox}
    \end{subtable}
    \vspace{2pt}

    \begin{minipage}{\linewidth}
        \raggedright
        \footnotesize
        $^{\ddagger}$ Chain rows count distinct shielding addresses within each chain; Total deduplicates identical EVM addresses across chains.
    \end{minipage}
\end{table}

\subsection{Commitment Set Sizes}\label{ssec:results_omega}

We evaluate the cumulative effects of \autoref{sec:deanonymization} on the Commitment Set Size $|\Omega(i)|$ defined in \autoref{ssec:privacy_metrics}. After cumulative pruning and cross-proof propagation, the mean size relative to the temporal baseline decreases by 82.2\%, 66.7\%, 58.5\%, and 38.2\% on Railgun's Ethereum, Arbitrum, Polygon, and BNB Chain deployments, respectively, and by 72.8\%, 76.7\%, 75.8\%, 77.5\%, and 80.4\% on Hinkal's Ethereum, Arbitrum, Polygon, Base, and Optimism pools. Public token constraints are a major, persistent filter in both protocols, and Value and propagation further refine the survivors. Railgun also benefits from Tree Number after a rollover, unlike Hinkal's long-lived single-tree pools. Hinkal's cumulative reductions are consequently more consistent across chains, whereas Railgun varies with tree count, asset composition, and history.

\begin{figure*}[t]
    \centering
    \begin{subfigure}[t]{\textwidth}
        \centering
        \includegraphics[width=\linewidth, trim=0 10 70 0, clip]{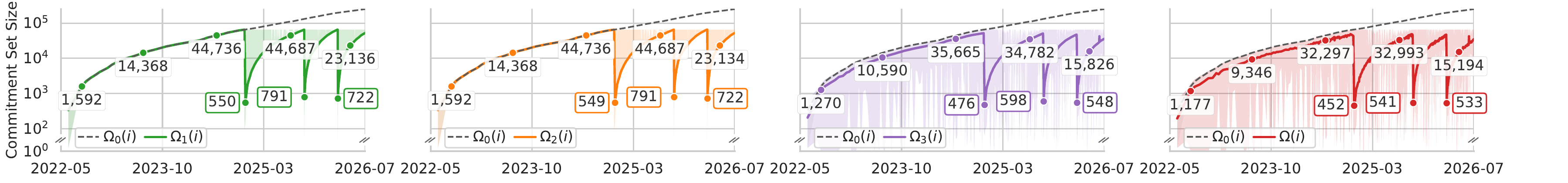}
        \caption{Ethereum}
    \end{subfigure}

    \vspace{0.4em}

    \begin{subfigure}[t]{\textwidth}
        \centering
        \includegraphics[width=\linewidth, trim=0 10 70 0, clip]{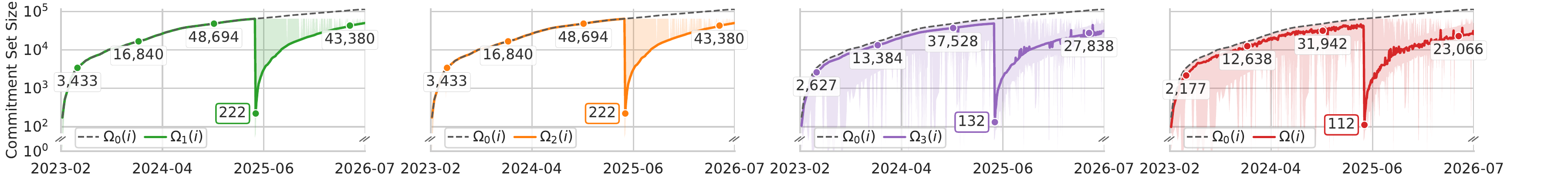}
        \caption{Arbitrum}
    \end{subfigure}

    \vspace{0.4em}

    \begin{subfigure}[t]{\textwidth}
        \centering
        \includegraphics[width=\linewidth, trim=0 10 70 0, clip]{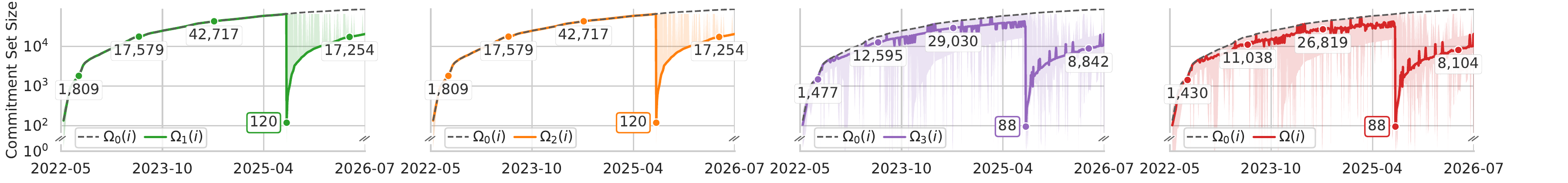}
        \caption{Polygon}
    \end{subfigure}

    \vspace{0.4em}

    \begin{subfigure}[t]{\textwidth}
        \centering
        \includegraphics[width=\linewidth, trim=0 10 70 0, clip]{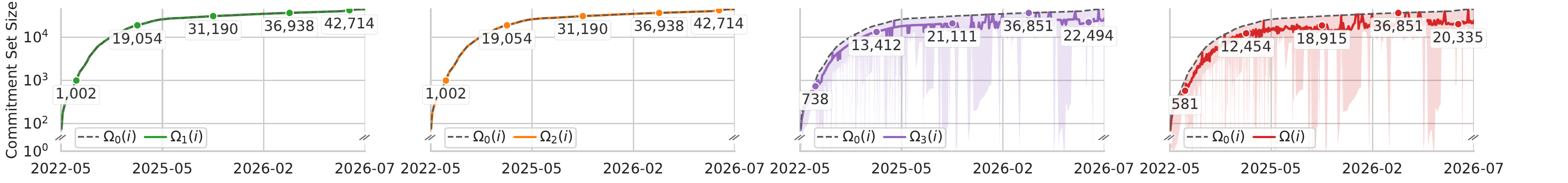}
        \caption{BNB Chain}
    \end{subfigure}

    \caption{Temporal evolution of Railgun Commitment Set Size through July~30,~2026 (UTC), in 500 equal-count block-order bins per chain. Successive panels compare $\Omega_0(i)$ with $\Omega_1(i)$, $\Omega_2(i)$, $\Omega_3(i)$, and the propagated final $\Omega(i)$ after Value. Curves show P50; shading spans P0--P100; labels mark selected times, with colored outlines identifying tree rollovers.}
    \label{fig:commitment_set_size_multichain}
\end{figure*}

\begin{figure}[t]
    \centering
    \includegraphics[width=0.98\linewidth,trim=5 110 10 20,clip]{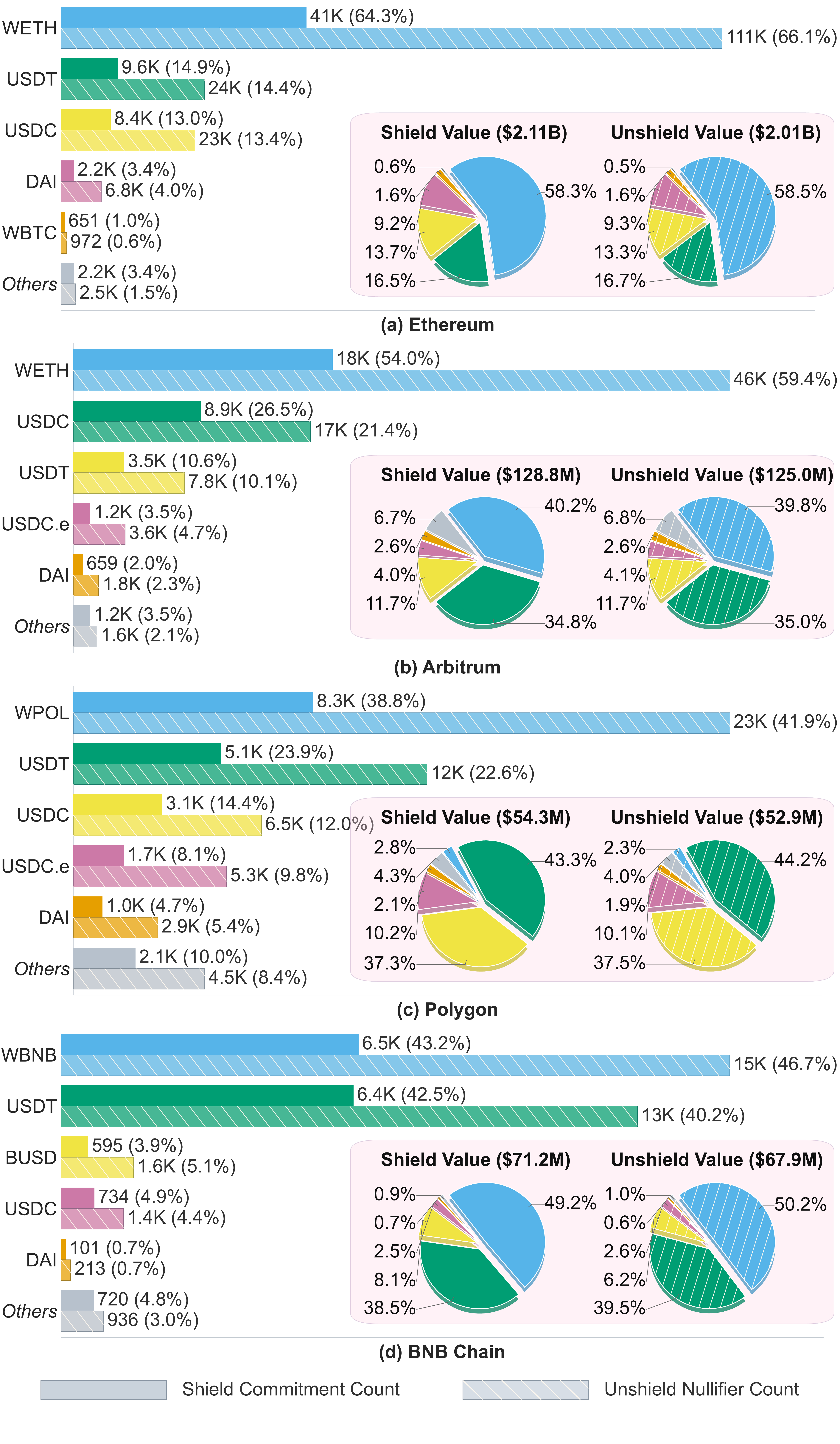}
    \caption{Top token activity across Railgun's four deployment chains. The inset pies show USD-value shares for priced tokens, computed as described in \appref{app:token-usd-aggregation}.}
    \label{fig:top_tokens_activity_multichain}
\end{figure}

\begin{figure}[t]
    \centering
    \begin{subfigure}[t]{0.515\linewidth}
        \centering
        \includegraphics[width=\linewidth,trim=5 5 40 0, clip]{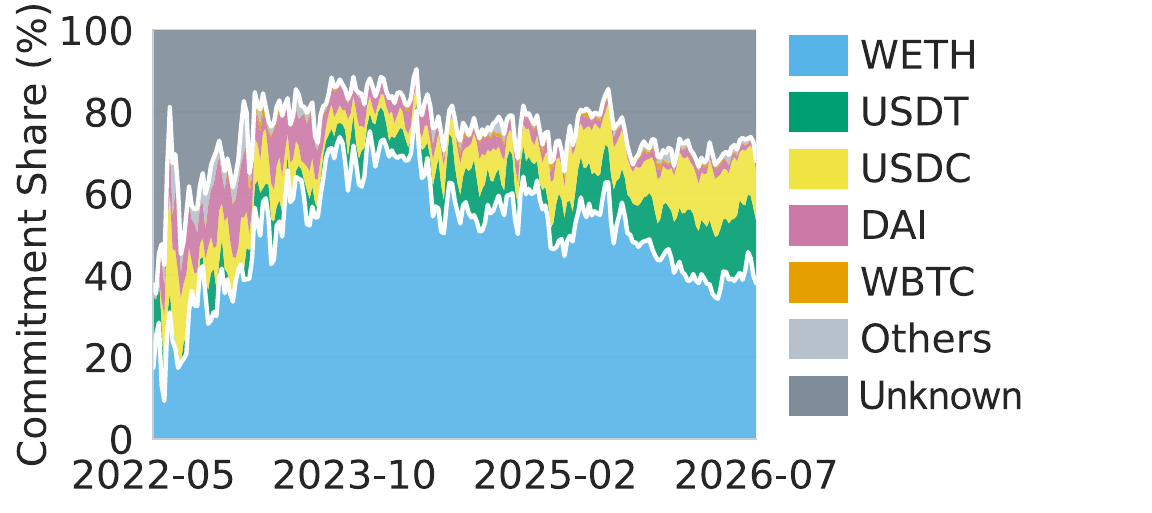}
        \caption{Ethereum}
    \end{subfigure}\hfill
    \begin{subfigure}[t]{0.485\linewidth}
        \centering
        \includegraphics[width=\linewidth, trim=30 5 40 0, clip]{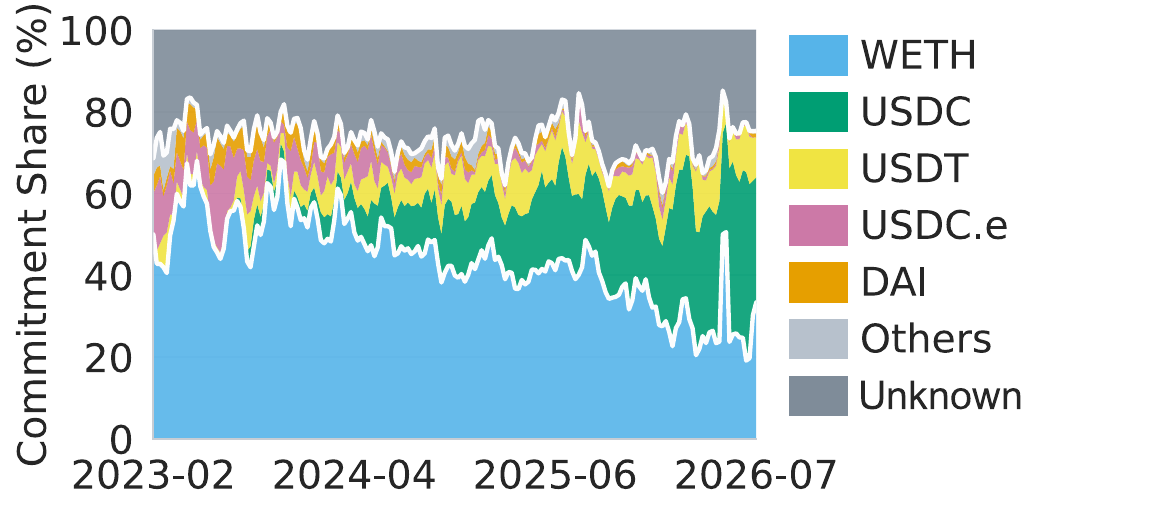}
        \caption{Arbitrum}
    \end{subfigure}

    \vspace{0.35em}

    \begin{subfigure}[t]{0.515\linewidth}
        \centering
        \includegraphics[width=\linewidth, trim=5 5 40 0, clip]{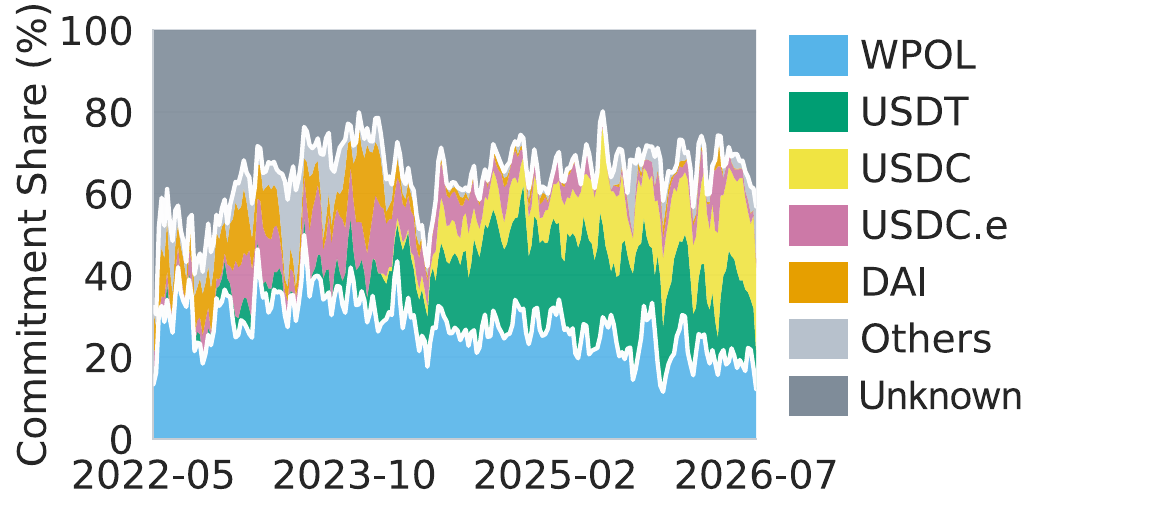}
        \caption{Polygon}
    \end{subfigure}\hfill
    \begin{subfigure}[t]{0.485\linewidth}
        \centering
        \includegraphics[width=\linewidth, trim=30 5 40 0, clip]{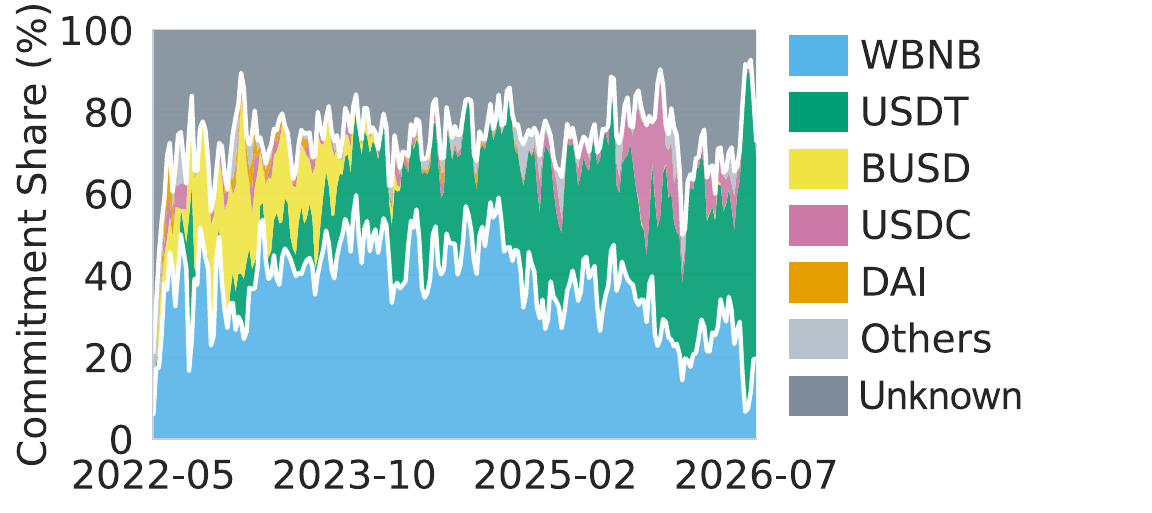}
        \caption{BNB Chain}
    \end{subfigure}

    \caption{Token shares of all Railgun commitments over time.}
    \label{fig:token_commitment_share_multichain}

\end{figure}

\point{Tree Number}\label{ssec:r1}
Railgun distributes commitments across multiple trees, each with capacity $2^{16}=65{,}536$. Because a proof's tree number is public and its Merkle root authenticates only that tree's state, $\Omega_1(i)$ contains at most the commitments already inserted there. Tree Number can therefore sharply reduce the candidate set immediately after a new tree opens. In \autoref{fig:commitment_set_size_multichain}, the local median Commitment Set Size falls to 550, 791, and 722 after Ethereum's three rollovers, and to 222 and 120 after the rollovers on Arbitrum and Polygon, respectively. BNB Chain remains in one tree throughout the observation window and shows no such reduction. By contrast, \autoref{tab:hinkal_deployments} shows that each selected Hinkal pool maintains one long-lived tree of height~200, with capacity $2^{199}$ leaves. These pools require no Railgun-style rollovers within the observation window and thus have no Tree Number stage.

\point{Proof Root}\label{ssec:r2}
Within the tree selected by Tree Number, or within an independent Hinkal pool, Proof Root further excludes commitments not covered by the authenticated state used by the proof. Its aggregate effect is limited because most proofs use a root close to the latest state available at execution time; nevertheless, a proof using an older root can exclude a substantial suffix of later commitments. In the Polygon case examined in \autoref{tab:case_examples}, Proof Root reduces the candidate set from 1{,}017 to 491. Across the five selected Hinkal pools, this stage reduces each pool's mean Commitment Set Size by less than 0.03\%, indicating that proofs generally use roots close to the latest available state.

\point{Token}\label{ssec:r3}
Token further excludes, from the Proof Root output, commitments incompatible with the target proof's public token information. In Railgun, proof-created commitments may inherit multiple feasible token labels: \autoref{fig:top_tokens_activity_multichain} shows which tokens commonly appear in shield and unshield activity, while \autoref{fig:token_commitment_share_multichain} shows their distribution in the historical commitment pool; together, these distributions determine how broad a target token branch remains. When WETH dominates both activity and candidate history, the common branches on Ethereum and Arbitrum remain broad; Polygon's more dispersed token composition, and BNB Chain's split between the two major WBNB and USDT branches, allow more commitments in other branches to be excluded once a token is fixed. Measured over all proof relations on each chain, the cumulative decrease in mean Commitment Set Size from $\Omega_0(i)$ through $\Omega_3(i)$ is approximately 79.1\%, 62.1\%, 55.7\%, and 34.3\% on Ethereum, Arbitrum, Polygon, and BNB Chain, respectively. By contrast, Hinkal has no uncertainty from inherited token labels: every same-token input--output relation publicly exposes an exact token address, so its candidates are tested against a singleton token branch and asset composition affects only the size of that branch. As shown in \autoref{fig:hinkal_commitment_set_size_multichain}, the cumulative decreases on Ethereum, Arbitrum, Polygon, Base, and Optimism are 69.6\%, 72.0\%, 72.5\%, 71.8\%, and 76.5\%, respectively, reflecting the strong partitioning effect of public token information.

\point{Value}\label{ssec:r4}
Value uses public amounts and balance relations to test feasibility within the branches that survive Token. After Value and propagation, the mean final Commitment Set Size ranges from 13{,}736 to 22{,}286 across the four Railgun chains and from 861 to 2{,}716 across the five selected Hinkal pools. Token remains the main persistent filter, while the final checkpoint removes candidates ruled out by numerical feasibility and its propagated consequences. Hinkal's smaller final sets also reflect its smaller independent pools and exact token addresses, which remove cross-token candidates before Value. The figures report $\Omega_1(i)$, $\Omega_2(i)$, and $\Omega_3(i)$ after the first three cumulative stages and the propagated final set $\Omega(i)$ after Value; $\Omega_4(i)$ before propagation is not reported separately. \autoref{ssec:note_source_pruning} explains the order and representation dependencies.

\subsection{Anonymity Set Sizes}\label{ssec:results_as}

\begin{figure}[t]
    \centering
    \begin{subfigure}[t]{0.512\columnwidth}
        \centering
        \includegraphics[width=\linewidth, trim=0 10 10 0, clip]{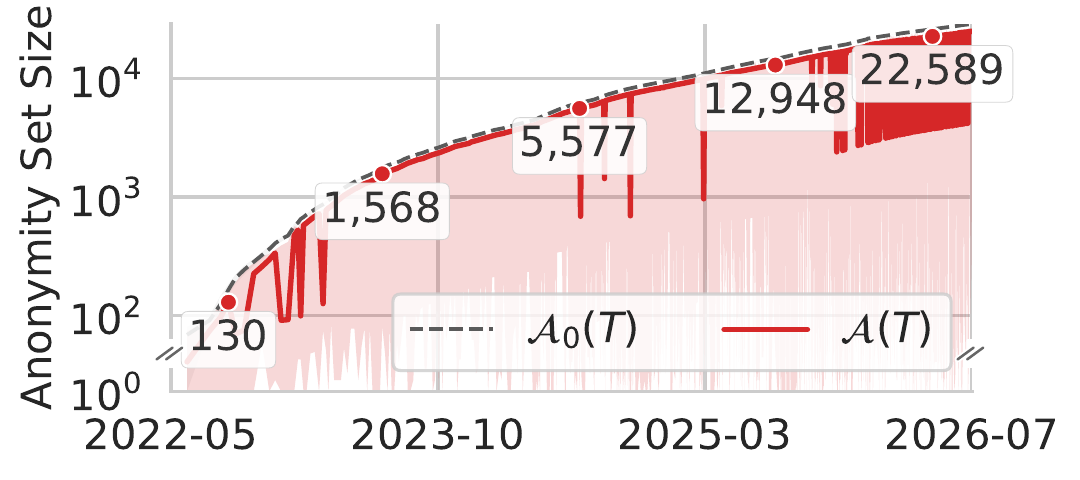}
        \caption{Ethereum}
    \end{subfigure}\hfill
    \begin{subfigure}[t]{0.485\columnwidth}
        \centering
        \includegraphics[width=\linewidth, trim=35 10 10 0, clip]{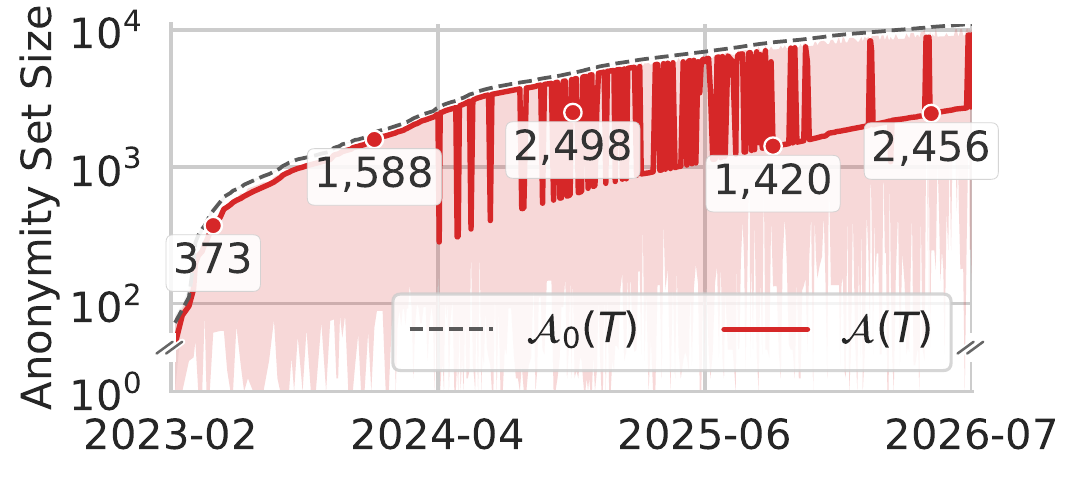}
        \caption{Arbitrum}
    \end{subfigure}

    \vspace{0.4em}

    \begin{subfigure}[t]{0.512\columnwidth}
        \centering
        \includegraphics[width=\linewidth, trim=0 10 10 0, clip]{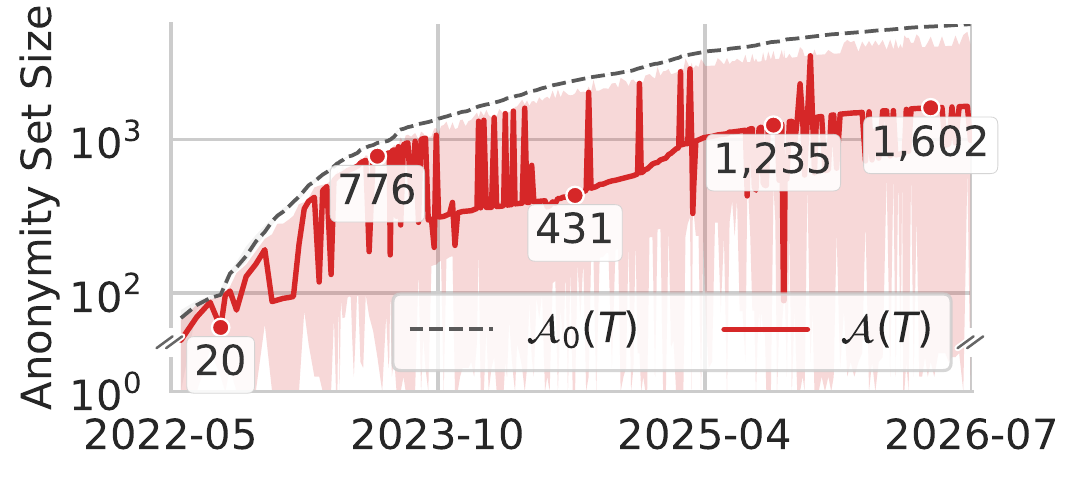}
        \caption{Polygon}
    \end{subfigure}\hfill
    \begin{subfigure}[t]{0.485\columnwidth}
        \centering
        \includegraphics[width=\linewidth, trim=35 10 10 0, clip]{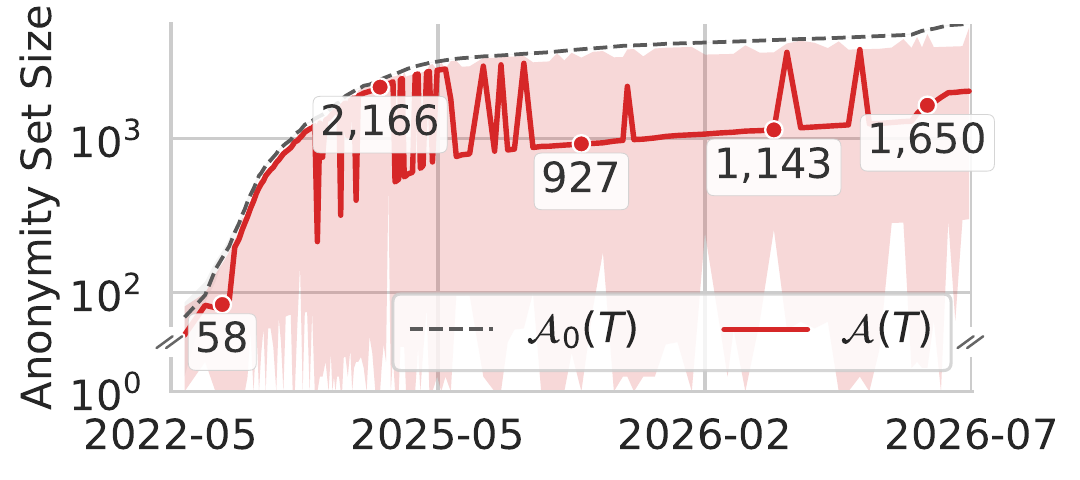}
        \caption{BNB Chain}
    \end{subfigure}

    \caption{Railgun Anonymity Set Size over time: temporal baseline (dashed) and final transaction-level union after closure lifting (solid). P50 curves use equal-count block-order bins of about 100 transactions; P0--P100 shading and labels at the 5\%, 25\%, 50\%, 75\%, and 95\% timeline positions.}
    \label{fig:anonymity_set_size_multichain}
\end{figure}

\begin{table}[t]
\centering
\tiny
\setlength{\tabcolsep}{1.8pt}
\renewcommand{\arraystretch}{1.08}
\caption{Final Anonymity Set Size by protocol and chain; buckets give transaction counts and row percentages.}
\label{tab:as_bucket_summary}
\begin{subtable}{\linewidth}
\caption{Railgun.}
\label{tab:railgun_as_bucket_summary}
\centering
\resizebox{\linewidth}{!}{%
\begin{tabular}{@{}lccccccccc@{}}
\toprule
\multirow{2}{*}{Chain} & \multicolumn{5}{c}{Count (\%)} & \multirow{2}{*}{Mean} & \multirow{2}{*}{Median} & \multirow{2}{*}{Max} & \multirow{2}{*}{Min} \\
\cmidrule(lr){2-6}
& 1 & 2--10 & 11--100 & 101--1000 & 1001+ & & & \\
\midrule
Ethereum & 373 (0.4\%) & 669 (0.8\%) & 1{,}477 (1.7\%) & 10{,}667 (12.5\%) & 71{,}971 (84.5\%) & 8{,}796 & 5{,}881 & 26{,}509 & 1 \\
Arbitrum & 153 (0.4\%) & 411 (1.0\%) & 1{,}503 (3.6\%) & 11{,}083 (26.2\%) & 29{,}106 (68.9\%) & 3{,}066 & 1{,}993 & 10{,}305 & 1 \\
Polygon & 243 (0.9\%) & 440 (1.6\%) & 2{,}194 (7.9\%) & 14{,}714 (53.1\%) & 10{,}128 (36.5\%) & 1{,}058 & 646 & 4{,}998 & 1 \\
BNB Chain & 151 (1.0\%) & 528 (3.6\%) & 1{,}003 (6.8\%) & 5{,}751 (39.1\%) & 7{,}275 (49.5\%) & 1{,}297 & 987 & 5{,}273 & 1 \\
\midrule
Total & 920 (0.5\%) & 2{,}048 (1.2\%) & 6{,}177 (3.6\%) & 42{,}215 (24.9\%) & 118{,}480 (69.8\%) & 5{,}458 & 2{,}445 & 26{,}509 & 1 \\
\bottomrule
\end{tabular}%
}
\end{subtable}

\vspace{5pt}

\begin{subtable}{\linewidth}
\caption{Hinkal.}
\label{tab:hinkal_as_bucket_summary}
\centering
\resizebox{\linewidth}{!}{%
\begin{tabular}{@{}lccccccccc@{}}
\toprule
\multirow{2}{*}{Chain} & \multicolumn{5}{c}{Count (\%)} & \multirow{2}{*}{Mean} & \multirow{2}{*}{Median} & \multirow{2}{*}{Max} & \multirow{2}{*}{Min} \\
\cmidrule(lr){2-6}
& 1 & 2--10 & 11--100 & 101--1000 & 1001+ & & & \\
\midrule
Ethereum & 44 (2.3\%) & 45 (2.4\%) & 338 (17.9\%) & 1{,}457 (77.3\%) & 0 (0.0\%) & 202 & 198 & 702 & 1 \\
Arbitrum & 27 (1.7\%) & 79 (4.9\%) & 470 (29.1\%) & 1{,}037 (64.3\%) & 0 (0.0\%) & 140 & 137 & 608 & 1 \\
Polygon & 56 (1.8\%) & 75 (2.4\%) & 885 (28.0\%) & 2{,}140 (67.8\%) & 0 (0.0\%) & 156 & 155 & 625 & 1 \\
Base & 138 (2.4\%) & 95 (1.6\%) & 743 (12.7\%) & 4{,}309 (73.6\%) & 566 (9.7\%) & 499 & 433 & 2{,}181 & 1 \\
Optimism & 43 (1.1\%) & 109 (2.7\%) & 1{,}508 (37.6\%) & 2{,}352 (58.6\%) & 0 (0.0\%) & 119 & 117 & 410 & 1 \\
\midrule
Total & 308 (1.9\%) & 403 (2.4\%) & 3{,}944 (23.9\%) & 11{,}295 (68.4\%) & 566 (3.4\%) & 272 & 171 & 2{,}181 & 1 \\
\bottomrule
\end{tabular}%
}
\end{subtable}
\end{table}

This subsection first answers the source-range part of the research question in \autoref{sec:introduction}: \emph{from which shielding addresses could the funds released by a transaction that executes unshield have originated?} \autoref{tab:as_bucket_summary} reports the final feasible-address distributions. Among 169{,}840 Railgun unshielding spend transactions, 2{,}968 retain at most ten addresses, including 920 singletons; however, 118{,}480 (69.8\%) still retain more than 1{,}000. Ethereum has the largest mean final set, at 8{,}796, whereas Polygon has the smallest, at 1{,}058. Among Hinkal's 16{,}516 unshielding spend transactions, 711 retain at most ten addresses, including 308 singletons, and the chain-level mean final sets range from 119 to 499.\footnote{Unlike Railgun, Hinkal permits relayed Shields. When Hinkal releases funds publicly, recipient transfers and relayer fees both appear as pool outflows. We therefore identify shielding addresses and exclude pure relayer-fee branches using contract semantics and asset transfers; see \appref{app:protocol-instantiations}.} Hinkal's smaller absolute sets partly reflect the shorter histories and smaller temporal baselines of its independent pools; a cross-protocol comparison must therefore also consider reduction relative to each protocol's own baseline.

The second part of the research question asks how far public protocol constraints can reduce this address set. \autoref{tab:as_summary} shows mean Anonymity Set Size reductions of 40.1\%, 46.3\%, 59.0\%, and 47.6\% for Railgun on Ethereum, Arbitrum, Polygon, and BNB Chain, respectively, and 57.0\%, 57.4\%, 53.0\%, 52.0\%, and 48.9\% for Hinkal on Ethereum, Arbitrum, Polygon, Base, and Optimism. The Railgun chronology in \autoref{fig:anonymity_set_size_multichain}, together with the corresponding Hinkal chronology in \appref{app:supporting_anonymity}, shows that these reductions persist throughout the observation window rather than arising from a few extreme transactions. For Railgun, much of the cross-chain variation reflects token-branch width. As \autoref{fig:top_tokens_activity_multichain} and \autoref{fig:token_commitment_share_multichain} jointly show, public unshield activity determines which token branches are queried frequently, while the asset distribution of historical commitments determines how wide those branches are. WETH dominates both activity and historical candidates on Ethereum and Arbitrum, so common transactions still fall into broad branches. Polygon's more dispersed asset distribution generally allows a fixed target token to exclude a larger share of historical commitments, explaining its stronger address-level reduction. Hinkal exposes the exact asset address for every same-token relation and therefore creates a sharper partition at the commitment layer. After recursive closure, however, each surviving commitment may still connect to many historical shielding addresses, so this local advantage does not yield a proportional final address-set reduction.

Proof composition further affects the final result. \autoref{tab:as_summary} shows that Railgun Proof~\fcirc{2}-only transactions have mean reductions of 53.4\%--74.1\%, higher than the 38.4\%--56.7\% for Proof~\fcirc{4}-only transactions; the latter release funds publicly while continuing to create shielded commitments and therefore generally retain broader feasible histories. Hinkal shows no consistent ordering between the two proof types, whose chain-level means differ by only 0.9--4.1 percentage points. Both protocols instead show a consistent within-transaction union effect: because the final set is the deduplicated union of proof-level address sets, one broad proof can offset substantial pruning in the others, so final privacy depends on their widths and overlaps rather than the most strongly pruned proof.

Recursive depth also relates to final Anonymity Set Size (\autoref{tab:closure_depth_summary}). We count one hop whenever backward tracing crosses one historical private state transition and record the maximum over all feasible source paths per transaction. The chain-level medians range from 4{,}628.5 to 24{,}872 for Railgun, substantially above Hinkal's 645--2{,}608, showing that Railgun's feasible sources usually extend through deeper histories. To isolate within-transaction union, we retain transactions for which exactly one proof contributes to $\mathcal{A}(T)$, rank them by maximum hops within each chain, and compare the bottom and top quartiles. Railgun's mean reduction falls from 54.5\%--66.0\% in the shallowest quartile to 12.3\%--43.6\% in the deepest; Hinkal's corresponding ranges are 61.6\%--73.5\% and 45.6\%--55.3\%. Deeper feasible ancestry therefore generally accumulates more historical shielding addresses during closure lifting and weakens final reduction.

\subsection{Patterns and Case Studies}\label{ssec:results_patterns}

\begin{table*}[t]
    \centering
    \tiny
    \newcommand{\tabtypesym}[1]{\raisebox{-0.18ex}{\scalebox{0.68}{\fcirc{#1}}}}
    \newcommand{\txlink}[2]{\href{#1}{\texttt{#2}}}
    \newcommand{\timelink}[2]{\href{#1}{#2}}
    \setlength{\tabcolsep}{1.5pt}
    \renewcommand{\arraystretch}{1.08}
    \caption{Representative Railgun unshielding spends illustrating the anonymity-loss patterns.}
    \label{tab:case_examples}
    \resizebox{\textwidth}{!}{%
        \begin{tabular}{c l p{1.95cm} c c l r r r r r r r r r r}
            \toprule
            \textbf{Pattern ID}  & \textbf{Chain}             & \textbf{Spend Tx}                                                                                                                     & \textbf{Time (UTC)}                                                                                                                      & \textbf{Proof} & \textbf{Token}  & \multicolumn{1}{c}{\textbf{Value}} & \textbf{$|\Omega_0(i)|$} & \textbf{$|\Omega_1(i)|$} & \textbf{$|\Omega_2(i)|$} & \textbf{$|\Omega_3(i)|$} & \textbf{$|\Omega(i)|$} & \textbf{$|\mathcal{E}(i)|$} & \textbf{$|\mathcal{A}_0(T)|$} & \textbf{$|\mathcal{A}(T)|$} & \textbf{$\mathrm{red}_{\mathcal{A}}(T)$} \\
            \midrule
            1.1                  & Arbitrum                   & \txlink{https://arbiscan.io/tx/0x16f6d4eca7cd630b9890f9601e76f493967292bd7633f6a7b943a6eeff4908f4}{0x16f6d4...08f4}                   & \timelink{https://arbiscan.io/tx/0x16f6d4eca7cd630b9890f9601e76f493967292bd7633f6a7b943a6eeff4908f4}{2025-05-08 11:44}                   & \tabtypesym{4} & USDC            & 3{,}480                            & 65{,}548                 & 13                       & 13                       & 8                        & 7                      & 3{,}186                               & 6{,}602                       & 1{,}014                     & 84.64\%                                   \\
            \midrule
            1.2                  & Polygon                    & \txlink{https://polygonscan.com/tx/0x30c4594c22ac0748cc26cf659f13f4a0d2d04d639685ad0acb1324d01c3a2259}{0x30c459...2259}               & \timelink{https://polygonscan.com/tx/0x30c4594c22ac0748cc26cf659f13f4a0d2d04d639685ad0acb1324d01c3a2259}{2022-07-01 18:14}               & \tabtypesym{4} & WETH            & 0.997436                           & 1{,}017                  & 1{,}017                  & 491                      & 182                      & 177                    & 6                                     & 60                            & 6                           & 90.00\%                                   \\
            \midrule
            1.3                  & Ethereum                   & \txlink{https://etherscan.io/tx/0x54ae52bfb2990c52715240a6a2485b8ab11e02e0081702cc909d93e96034b41b}{0x54ae52...b41b}                  & \timelink{https://etherscan.io/tx/0x54ae52bfb2990c52715240a6a2485b8ab11e02e0081702cc909d93e96034b41b}{2024-11-18 04:41}                  & \tabtypesym{4} & AAVE            & 100.250627                         & 63{,}213                 & 63{,}213                 & 63{,}213                 & 46                       & 46                     & 1                                     & 8{,}766                       & 1                           & 99.99\%                                   \\
            1.3                  & Ethereum                   & \txlink{https://etherscan.io/tx/0x8a77c423fc2f3c5972f49cd0bd15c6c73cb63fbdb693af81df543dd346258a3a}{0x8a77c4...8a3a}                  & \timelink{https://etherscan.io/tx/0x8a77c423fc2f3c5972f49cd0bd15c6c73cb63fbdb693af81df543dd346258a3a}{2024-01-24 06:27}                  & \tabtypesym{2} & ENS (\acs{NFT}) & 1                                  & 26{,}477                 & 26{,}477                 & 26{,}477                 & 2{,}015                  & 2{,}015                & 1                                     & 3{,}552                       & 1                           & 99.97\%                                   \\
            \midrule
            1.4                  & BNB Chain                  & \txlink{https://bscscan.com/tx/0xf70917dd9d42531db85dc815508c16ead5d3d74d34067b3351c3fb793f0e9471}{0xf70917...9471}                   & \timelink{https://bscscan.com/tx/0xf70917dd9d42531db85dc815508c16ead5d3d74d34067b3351c3fb793f0e9471}{2022-08-19 14:50}                   & \tabtypesym{4} & WBNB            & 183.064222                         & 407                      & 407                      & 407                      & 304                      & 97                     & 61                                    & 42                            & 32                          & 23.81\%                                   \\
            \midrule
            \multirow{2}{*}{2.1} & Ethereum                   & \txlink{https://etherscan.io/tx/0xb0d961aa8cb4c93bce7543908e6fbfb5282fd6bd6e0d047c52e11bcba8e3d7df}{0xb0d961...d7df}                  & \timelink{https://etherscan.io/tx/0xb0d961aa8cb4c93bce7543908e6fbfb5282fd6bd6e0d047c52e11bcba8e3d7df}{2025-01-29 23:09}                  & \tabtypesym{2} & WETH            & 332.787559                         & 73{,}045                 & 65{,}536                 & 65{,}536                 & 52{,}158                 & 38{,}929               & 13{,}165                              & 10{,}142                      & 8{,}155                     & 19.59\%                                   \\
                                 & Ethereum                   & \txlink{https://etherscan.io/tx/0x8ebf10b7523b9efc2918875f022657ac02dca953cca4ac39a24167dbefc98342}{0x8ebf10...8342}                  & \timelink{https://etherscan.io/tx/0x8ebf10b7523b9efc2918875f022657ac02dca953cca4ac39a24167dbefc98342}{2024-04-16 15:57}                  & \tabtypesym{4} & WETH            & 0.512282                           & 31{,}815                 & 31{,}815                 & 31{,}815                 & 25{,}309                 & 25{,}309               & 6{,}411                               & 4{,}406                       & 3{,}995                     & 9.33\%                                    \\
            \midrule
            \multirow{2}{*}{2.2} & \multirow{2}{*}{Arbitrum}  & \multirow{2}{*}{\txlink{https://arbiscan.io/tx/0xc0ea773f5b7a4880f7390b300beed3a146200622cedf629ee401720357d8f050}{0xc0ea77...f050}}  & \multirow{2}{*}{\timelink{https://arbiscan.io/tx/0xc0ea773f5b7a4880f7390b300beed3a146200622cedf629ee401720357d8f050}{2024-01-05 12:43}}  & \tabtypesym{4} & USDC            & 9{,}950                            & 16{,}372                 & 16{,}372                 & 16{,}372                 & 2{,}589                  & 2{,}397                & 161                                   & \multirow{2}{*}{1{,}712}      & \multirow{2}{*}{98}         & \multirow{2}{*}{94.28\%}                  \\
                                 &                            &                                                                                                                                       &                                                                                                                                          & \tabtypesym{4} & WBTC            & 0.17                               & 16{,}372                 & 16{,}372                 & 16{,}372                 & 3{,}990                  & 3{,}970                & 34                                    &                               &                             &                                           \\
            \midrule
            \multirow{2}{*}{3.1} & \multirow{2}{*}{Ethereum}  & \multirow{2}{*}{\txlink{https://etherscan.io/tx/0x520db58d2c52a97617ec418fd58cf9d7590bcdc5afb870b9bcdd46a4ae117820}{0x520db5...7820}} & \multirow{2}{*}{\timelink{https://etherscan.io/tx/0x520db58d2c52a97617ec418fd58cf9d7590bcdc5afb870b9bcdd46a4ae117820}{2025-02-24 10:30}} & \tabtypesym{4} & WETH            & 0.048337                           & 77{,}981                 & 12{,}445                 & 12{,}445                 & 10{,}173                 & 9{,}912                & 15{,}553                              & \multirow{2}{*}{10{,}694}     & \multirow{2}{*}{9{,}620}    & \multirow{2}{*}{10.04\%}                  \\
                                 &                            &                                                                                                                                       &                                                                                                                                          & \tabtypesym{2} & FXI             & 50{,}000                           & 77{,}981                 & 12{,}445                 & 12{,}445                 & 8                        & 7                      & 1                                     &                               &                             &                                           \\
            \midrule
            \multirow{3}{*}{3.2} & \multirow{3}{*}{BNB Chain} & \multirow{3}{*}{\txlink{https://bscscan.com/tx/0x78d080aae47e518c405a4fccdd6c324ae3f11817572c30dd873adca549962ca7}{0x78d080...2ca7}}  & \multirow{3}{*}{\timelink{https://bscscan.com/tx/0x78d080aae47e518c405a4fccdd6c324ae3f11817572c30dd873adca549962ca7}{2023-07-25 08:32}}  & \tabtypesym{2} & USDC            & 34.445306                          & 7{,}062                  & 7{,}062                  & 7{,}062                  & 2{,}065                  & 2{,}027                & 52                                    & \multirow{3}{*}{800}          & \multirow{3}{*}{30}         & \multirow{3}{*}{96.25\%}                  \\
                                 &                            &                                                                                                                                       &                                                                                                                                          & \tabtypesym{2} & USDC            & 51.1101                            & 7{,}062                  & 7{,}062                  & 7{,}062                  & 2{,}065                  & 2{,}029                & 52                                    &                               &                             &                                           \\
                                 &                            &                                                                                                                                       &                                                                                                                                          & \tabtypesym{2} & USDC            & 11.657887                          & 7{,}062                  & 7{,}062                  & 7{,}062                  & 2{,}065                  & 1{,}975                & 52                                    &                               &                             &                                           \\
            \bottomrule
        \end{tabular}%
    }
\end{table*}

To expose the mechanisms behind \autoref{ssec:results_as}, \autoref{tab:case_examples} traces representative Railgun unshielding spend transactions from $\Omega_0(i)$ through cumulative pruning and cross-proof propagation to $\Omega(i)$, then through closure lifting to $\mathcal{E}(i)$ and $(|\mathcal{A}_0(T)|,|\mathcal{A}(T)|)$. The cases cover proof-level specialization, proof-structure effects, and transaction-level aggregation.

\point{Proof-Level Specialization}
This category arises directly from the local pruning constraints of \autoref{ssec:note_source_pruning}. \emph{Pattern~1.1, tree-number fragmentation} appears in the Arbitrum rollover case \texttt{0x16f6d4...08f4}. The proof is in tree~1, whereas its temporal baseline still includes the complete history of tree~0, so Tree Number cuts the candidate set from 65{,}548 to 13. After the remaining stages and closure lifting, the final Anonymity Set contains 1{,}014 shielding addresses against a temporal baseline of 6{,}602. \emph{Pattern~1.2, proof-root staleness} captures the corresponding restriction within one tree. In the Polygon case \texttt{0x30c459...2259}, Tree Number is inactive, but Proof Root cuts the candidate set from 1{,}017 to 491. The remaining pipeline reduces the final Commitment Set to 177, and closure lifting leaves only six feasible shielding addresses from a baseline of 60. \emph{Pattern~1.3, asset specialization} is illustrated by the Ethereum AAVE case \texttt{0x54ae52...b41b} and the Ethereum ENS \acs{NFT} case \texttt{0x8a77c4...8a3a}. In the AAVE case, the structural stages do nothing, but Token cuts the candidate set from 63{,}213 to 46 and closure lifting leaves a singleton Anonymity Set. In the ENS case, the token branch still contains 2{,}015 commitments, yet closure lifting maps them to a single feasible shielding address. Thus, a narrow asset branch need not become a singleton at the commitment layer to converge to a very small address set. \emph{Pattern~1.4, value specialization} is illustrated by the BNB Chain WBNB case \texttt{0xf70917...9471}. Because WBNB retains a relatively broad historical token branch, Token narrows it only from 407 to 304; after Value and propagation, the final Commitment Set contains 97 candidates. Closure lifting reaches 61 historical shield transactions and 32 of the 42 baseline shielding addresses, a 23.81\% reduction. This case shows that value feasibility and propagation can add useful constraints even when the target asset's historical branch remains broad.

\point{Proof-Structure Effects}
The public input and output structure of a proof changes the strength of the Value constraint, helping to explain the aggregate difference between the Proof~\fcirc{2}-only and Proof~\fcirc{4}-only rows in \autoref{tab:as_summary}. \emph{Pattern~2.1, low-input rigidity} is illustrated by two Ethereum WETH cases. Transaction \texttt{0xb0d961...d7df} contains a Proof~\fcirc{2} with $m_i=1$ and $k_i=0$, so its single input must explain the complete public output value; after Value and propagation, the final Commitment Set falls from 52{,}158 to 38{,}929 candidates. By contrast, \texttt{0x8ebf10...8342} contains a Proof~\fcirc{4} with $m_i=2$ and $k_i=1$, for which the unreleased remainder may enter the newly created commitment; its candidate set stops shrinking after Token at 25{,}309. This local distinction does not by itself determine the transaction-level result, because the projected sets of multiple contributing proofs are subsequently unioned. \emph{Pattern~2.2, low-output rigidity} is the corresponding effect of $k_i$ in Proof~\fcirc{4}. When $k_i=1$, only one newly created shielded note can carry the unreleased remainder, narrowing the admissible total-input range in \autoref{ssec:s4}. In the Arbitrum case \texttt{0xc0ea77...f050}, both Proof~\fcirc{4} branches that produce public outputs have $k_i=1$. Their final Commitment Sets contain 2{,}397 and 3{,}970 candidates, and closure lifting maps them to 80 and 28 shielding addresses. The partially overlapping address sets yield a transaction-level union of 98 addresses, far below the temporal baseline of 1{,}712.

\point{Transaction-Level Aggregation}
At the transaction layer, $\mathcal{A}(T)$ is the deduplicated union of the shielding-address sets projected from all constituent proofs with public outputs, so the result depends on their widths and overlap. \emph{Pattern~3.1, broad-proof shielding} is the regime in which one broad proof dominates the union. In Ethereum transaction \texttt{0x520db5...7820}, the Proof~\fcirc{4} branch still maps to 9{,}620 feasible shielding addresses after closure, whereas the accompanying Proof~\fcirc{2} branch collapses to one. That singleton is already contained in the broader set, so the transaction-level union remains 9{,}620 and the overall reduction is only 10.04\%. A strongly pruned proof therefore cannot offset another broad proof in the same transaction. \emph{Pattern~3.2, overlap collapse} is the opposite regime. In BNB Chain transaction \texttt{0x78d080...2ca7}, the three Proof~\fcirc{2} branches have different final Commitment Sets, but closure lifting maps all three to the same 52 historical shield transactions and the same 30 shielding addresses. Because the projected sets coincide, deduplication adds no new addresses and the multi-proof transaction retains a final Anonymity Set Size of 30.

\point{Pattern Analysis of Small-Set Cases}
Small-set cases show that the final Anonymity Set Size depends not only on how many candidate commitments remain, but also on how many shielding addresses those commitments can reach under backward tracing. Tree Number and Proof Root restrict only which commitments the current proof may consume directly; they do not sever those commitments' links to earlier private states, because, for example, a commitment in a new tree may descend from funds represented by notes in an older tree. Accordingly, Pattern~1.1 leaves only seven commitments, yet closure lifting reaches 3{,}186 shield transactions and 1{,}014 shielding addresses; by contrast, the Ethereum ENS case retains 2{,}015 commitments, all of which trace to one shielding address. Token and Value instead restrict candidate branches using public asset and amount relations, and the token condition remains in force during backward tracing, so asset branches with smaller histories and more concentrated sources are more likely to yield small sets. At the transaction layer, any contributing proof that still maps to many shielding addresses can dominate the final union; a multi-proof transaction therefore forms a small set only when the projected address set of every contributing proof is narrow or when those sets overlap.

\section{Discussion}\label{sec:discussion}

\point{Performance--Anonymity Tradeoffs}
Railgun and Hinkal make different state-management tradeoffs (\appref{app:protocol-instantiations}). Railgun's fixed-depth-16 Merkle trees have predictable update paths, but a new tree opens when the current tree cannot accommodate the next commitment batch, and public tree numbers partition one deployment's history into candidate domains. Hinkal's much larger logical tree maintains only the depth reached by its state, so the evaluated pools neither roll over nor support Tree Number pruning. Railgun thus trades bounded per-tree state for structural partitioning, whereas Hinkal preserves continuity within each pool but lets the maintained path grow with the state. Tree height, capacity, and rollover policy jointly shape maintenance cost and how much history remains in one anonymity domain.

Root retention similarly balances state cost, proof-submission flexibility, and effective anonymity history. Railgun accepts historical roots indefinitely, keeping old proofs usable but growing retained on-chain state and allowing a stale root to fix the candidate set at an earlier prefix. Hinkal's circular history retains the latest 200 roots, bounding state growth and stale-root exposure; its mean Proof Root reduction remains below 0.03\%, at the cost of submitting a proof before its root expires or regenerating it afterward.

\point{Implications for Protocol Designers}
Upgrade policy likewise determines whether accumulated anonymity history survives across versions. On the three chains with a V1--V2 transition, Railgun continued the same production pool's commitment-tree state, so old commitments remain feasible sources in V2. Each Hinkal contract generation instead starts an independent pool, disconnecting its predecessors and excluding their commitments from new proofs. \autoref{tab:hinkal_deployments} shows that 23 of Hinkal's 37 nonempty pools contain fewer than 1{,}000 commitments; each new pool therefore restarts anonymity accumulation. Protocol upgrades should preserve pool state and anonymity-domain continuity where possible.

Public asset information additionally partitions each pool by token. Hinkal exposes the exact token address of each same-token relation, so Token reduces mean Commitment Set Size from the temporal baseline by 69.6\%--76.5\% across the five chains; Railgun shielded transfers do not directly reveal the exact token and can retain multiple feasible branches. This partition survives closure: in \appref{app:supporting_anonymity}, 1{,}813 of 1{,}884 Ethereum unshielding spend transactions (96.2\%) publicly release one token, and every feasible historical source in their complete closures has that token. Their mean shielding-address baseline falls from 485 to 198 when restricted to the public token, and the complete closure leaves 197. Hinkal's address-level reduction is milder than its commitment-level contraction because a transaction's Anonymity Set unions its constituent proofs' source sets, widening the transaction-level set but not any proof's token domain. Public token information thus causes persistent privacy loss across commitment, proof, and transaction layers.

\point{Implications for Users}
Users should use the freshest available root: an old Proof Root removes up to 10{,}384 candidates from a single Railgun proof in our data. After long delays or failed submissions followed by pool activity, users should update the witness and regenerate the proof rather than retry against old state. When compatible pools are available, they should prefer a mature active pool over a newer one with little history; publicly withdrawing from the old pool and Shielding into the new one creates a new public boundary without carrying over the old anonymity history. They should also judge the target asset's history, not total pool size. In \autoref{tab:case_examples}, Token reduces the AAVE case from 63{,}213 candidates to 46, after which closure reaches one shielding address; when flexible, users should let new or rarely used token branches accumulate history. In the WBNB case, Value and propagation reduce the 304 candidates surviving Token to 97; when possible, users should prefer common ranges to rare exact values.

\point{Toward Better Privacy Protocols}
Better privacy protocols should not make baseline anonymity depend on a particular time, asset, or pool. If an in-place upgrade cannot preserve the commitment-tree state, a new pool could verify a private migration proof against the old state, consume an old note, and create a new one without a public withdrawal and subsequent Shield, thereby preserving provenance. This mechanism must prevent cross-pool double-spending and address root finality and state synchronization. Likewise, token identity could remain inside commitments and zero-knowledge circuits, with the exact asset revealed only for necessary public settlement and token consistency and value conservation proved in-circuit. Private conversion proofs could keep a public output token from deterministically fixing its historical input token, at the cost of larger circuits and proofs and new liquidity, pricing, and manipulation-resistance constraints.

\section{Related Work}
\label{sec:related}

\point{Heuristic attribution-based privacy analysis}
Huseynov et al.\ apply five heuristics based on timing, address reuse, public-graph proximity, and amount patterns to Railgun on Ethereum, uniquely linking 17.65\% of withdrawals to one deposit and reporting a 3.42-bit median anonymity loss \cite{Huseynov2026Railgun}. Much of the broader privacy-measurement literature similarly approaches user-level privacy through clustering, timing, value patterns, or other heuristic attribution signals \cite{Meiklejohn2013Fistful,Androulaki2013BitcoinPrivacy,Yousaf2019CrossLedgerTracing,Kalodner2020BlockSci}, including studies of Zcash, mixers, Tornado Cash, and Umbra \cite{Kappos2018Zcash,Wang2023Mixers,Wu2022Tutela,Kovacs2024Umbra}. What is missing, especially for widely deployed shielded \acs{UTXO}-based protocols, is a non-heuristic baseline that first isolates anonymity loss from protocol-visible constraints alone and systematically traces recursive provenance through hidden-state transitions. We provide such a framework by decomposing the system across commitment, proof, and transaction layers and measuring address-level feasible sets without user attribution. This makes our results complementary to user-level analysis: entity linking, heuristic attribution, and probability-weighted metrics can be layered on top.

\point{Non-heuristic structure-based privacy analysis}
Ring-based systems provide the closest precedent for non-heuristic structural propagation: because each ring publishes its candidate members, eliminating one candidate can constrain subsequent transactions along the public graph. Monero studies combine such chain reactions with heuristic inference \cite{Kumar2017MoneroTrace,Moser2018MoneroTrace}, while later work formalizes structure-only attacks and studies graph-analysis resistance and sustainable ring construction \cite{Vijayakumaran2023DM,Egger2022GraphAnalysis,Chow2023SustainableRings}. Protocol-design research also spans commitment-nullifier payments \cite{BenSasson2014Zerocash,Rondelet2019ZETH}, account- and ledger-based confidentiality \cite{Bunz2020Zether,Cecchetti2017Solidus}, programmable private computation \cite{Kosba2016Hawk,Bowe2020Zexe,Steffen2019Zkay,Kerber2021Kachina}, and commitment-backed \acs{UTXO} stealth addresses \cite{StealthHub2025UTXO}; a recent systematization compares these designs \cite{Baldimtsi2024SoK}. These bodies of work establish structural inference techniques, protocol constructions, security notions, and programming models, but not a method for measuring recursive provenance in deployed shielded systems. The shielded \acs{UTXO}-based systems studied here present a more involved reconstruction problem: consumed commitments are not published, value can traverse multiple hidden-state transitions through newly created notes, and the relevant public constraints span the commitment, proof, and transaction layers. Analysis therefore cannot propagate over an already published candidate graph; it must first reconstruct each proof's feasible commitments, recursively lift them through historical hidden-state transitions to shielding addresses, and aggregate them at the transaction layer.

\section{Conclusion}\label{sec:conclusion}

This paper presents a non-heuristic privacy-measurement framework for shielded \acs{UTXO}-based protocols for \acs{DeFi}, grounded in a three-layer note/proof/transaction abstraction. Using only public ledger traces and protocol rules, it recursively traces each proof's feasible commitments through historical hidden-state transitions to shielding addresses and aggregates them into transaction-level Anonymity Set Size. Across the complete histories of all Railgun production deployments and five independent Hinkal pools, the framework yields mean Anonymity Set Size reductions of 40.1\%--59.0\% relative to the temporal baseline. The analysis deliberately excludes heuristic methods such as user/entity attribution, thereby isolating anonymity loss caused by public protocol constraints. The three-layer framework can also be instantiated, according to public protocol semantics, for broader commitment-nullifier privacy protocols beyond our target class: in Zcash, for example, structural tree/root constraints can still bound candidate history, while its single-asset design renders the Token stage inapplicable. Broader commitment-nullifier privacy protocols can thus first compare anonymity loss caused by public protocol constraints under a common non-heuristic baseline, then layer on heuristic analyses for specific applications, such as money-laundering prevention.

\appendix
\section*{Ethical Considerations}
\hypertarget{sec:ethics}{}

We considered the effects of this research on protocol users, developers and operators, and researchers or analysts. The study passively analyzes public, append-only blockchain data and public protocol semantics; it submits no transactions, changes no protocol state, uses no non-public data, and involves no human participants. Our analysis uses only protocol-visible constraints and makes no identity or ownership attributions, behavioral inferences, or probabilistic rankings.

The results are dual use: small or singleton feasible sets may help narrow transaction provenance, while the same measurements help users assess effective anonymity and designers mitigate privacy loss. We limit incremental harm by reporting feasible address sets rather than identities and by not combining results with off-chain attribution signals, although others could do so. Because the study measures public protocol behavior rather than exposing a secret implementation flaw, we do not treat it as requiring a separate vulnerability-disclosure process. We judged both the study and its publication justified because the traces and constraints are already public, while systematic analysis enables concrete defensive guidance and more privacy-preserving designs.

\section*{Open Science}
\hypertarget{sec:open-science}{}

An anonymized artifact is available at \url{https://anonymous.4open.science/r/the_anonymity_gap-BCDD/README.md}. It provides the complete acquisition-to-analysis workflow and code for reconstructing the public-chain inputs and reproducing commitment pruning, backward closure, transaction-level anonymity metrics, and closure-depth measurements for every evaluated deployment. Its linked data archive contains the exact prepared inputs and frozen outputs for the observation window ending July~30, 2026.

\bibliographystyle{unsrturl}
\bibliography{references_usenix}

\begingroup
\let\subsection\section
\subsection{Formal Propagation Procedure}\label{app:combined_propagation}

This appendix formalizes the propagation stage of \autoref{ssec:ss} by specifying its analysis order, state, update actions, and fixpoint. It uses the conservative local operator of \autoref{ssec:s4}: an update removes a candidate only when a public constraint or a mandatory spend makes it unavailable.

\point{Proof Relations and Public Order}
We use the proof abstraction $\pi_i$ defined in \autoref{ssec:proof}. It need not coincide one-to-one with a cryptographic proof: one cryptographic proof may jointly authenticate several such relations. Propagation requires an analysis order, not a claim that those relations were verified separately or sequentially.

Let $\prec$ denote the canonical order induced by the public state-transition trace. Across transactions, it follows ledger execution order; among relations within one transaction or jointly authenticated by one cryptographic proof, it uses only an order exposed by that trace. The order directs conservative information flow. In particular, commitments created within a spend transaction remain unavailable to every proof in that transaction, as required by \autoref{ssec:transaction}, and their projected labels can affect only later state transitions. If no internal order is public, the relations share one execution boundary and no order-dependent creating-proof edge is introduced.

\point{Propagated Proof State}
For each proof $\pi_i$, $\Xi(i)$ contains its fixed structural domain $\Omega_2(i)$, current Commitment Set $\Omega(i)$, mandatory set $M(i)$, token-guarded family $\Sigma(i)$, and projected output labels. The structural domain is produced by Strategies~1 and~2 and includes the identity case when either strategy is inapplicable. Every later state satisfies $M(i)\subseteq\Omega(i)\subseteq\Omega_2(i)$. Propagation never changes $\Omega_2(i)$; subsequent contraction comes from reservations or from reevaluating the token--value tail under narrower public interpretations.

The propagated state also retains the immutable public inputs introduced in \autoref{ssec:proof}, \autoref{ssec:s3}, and \autoref{ssec:s4}: $m_i$, $k_i$, $\mathcal{T}_i$, the public balance relation of $\pi_i$, and $U_i$ when $O_i\neq\varnothing$. They are fixed properties of $\pi_i$ and its public trace.

\point{Token-Guarded Representation}
The family $\Sigma(i)=\{\Omega^t(i)\}_{t\in\Theta(i)}$ records the surviving token interpretations of $\pi_i$, where $\Theta(i)\subseteq\Theta_i^\star$ is the current token-guard set. Each $\Omega^t(i)$ descends from the local slice $\Omega_4^t(i)$ returned by \autoref{ssec:s4}, and their union is the current Commitment Set, $\Omega(i)=\bigcup_{t\in\Theta(i)}\Omega^t(i)$. Whether $\Sigma(i)$ is a singleton is determined by the public trace, not by the Proof~\fcirc{2}--\fcirc{4} name. If the trace fixes one token, then $\Theta(i)=\{t\}$ and the unique guarded slice equals $\Omega(i)$. If several token interpretations remain possible, $\Sigma(i)$ retains one slice for each surviving interpretation. Removing a token branch removes its guard from $\Theta(i)$ and its slice from $\Sigma(i)$.

For each branch, $M^t(i)$ contains the commitments present in every explanation under guard $t$. The cross-branch set is $M(i)=\bigcap_{t\in\Theta(i)}M^t(i)$, so a commitment is globally mandatory only when every surviving token interpretation forces it.

\point{Projected Labels on Created Commitments}
We use the feasible token set $\Lambda(c)$ from \autoref{ssec:s3} and the creating proof $\pi(c)$ from \autoref{ssec:proof}. The current output state of $\pi(c)$ may further narrow $\Lambda(c)$ and the feasible value interval carried by $c$ under each surviving guard. Later proofs read these projected labels when reapplying $S_\mathrm{token}$ and $S_\mathrm{value}$.

The projection always respects the public creation relation. A token label is not copied blindly from an input branch: the same guard is inherited only when the public trace relates the created commitment to that token, and a different guard is propagated only when the trace determines the corresponding token relation. Likewise, a public token at creation yields a singleton label even if other relations jointly authenticated by the same cryptographic proof use different tokens. The value component is the current feasible output-value interval under each token guard, intersected with any value information exposed at creation.

The guarded proof state and the commitment labels are two views of the same semantic information. $\Sigma(i)$ stores the surviving interpretations of $\pi_i$; projected labels expose the corresponding output interpretation to later proofs that contain commitments created by $\pi_i$ in their structural domains. All output labels affected by an upstream branch removal or an output-interval contraction are synchronized before downstream recomputation. This forward semantic channel is separate from closure provenance. A proof-created commitment may depend on both a direct Shield source and earlier proof-created state. Propagation records its feasible token/value state; $\mathsf{Entry}$ and $\mathsf{Pred}$ in \autoref{ssec:closure_lifting} retain the publicly permitted provenance alternatives during backward lifting.

\point{Shared Propagation State}
Let $R$ be the shared reserved-commitment set. A commitment enters $R$ once it is mandatory for one proof. Because the corresponding note is consumed there, it is unavailable to any other compatible proof domain. This reservation uses only the commitment--nullifier single-spend rule of \autoref{ssec:note}; it does not select among merely optional candidates. The notification relation is instantiated conservatively from the public trace. Railgun removes a uniquely reserved commitment leaf from every other compatible domain whose authenticated root covers it. Hinkal processes reservations chronologically, reserving all commitments for a proof relation only when its surviving domain equals its nonzero input count, and suppresses this rule when duplicate commitment values prevent a unique mapping to leaf occurrences. Both instantiations remove only commitments whose reuse is impossible; their scheduling affects pruning completeness, not soundness.

Let $Q$ be a fair work queue of proofs whose semantic states may shrink. A proof is enqueued when either a mandatory set enlarges $R$ or a projected label on one of its proof-created candidates narrows. These are the two propagation channels: the reservation channel removes notes forced into another spend, while the label channel propagates narrowed token/value labels through created commitments.

\begin{figure}[t]
    \centering
    \footnotesize
    \begin{tcolorbox}[
            enhanced,
            title=Cross-Proof Propagation $\Pi_\mathrm{prop}$,
            attach boxed title to top left={xshift=2mm,yshift=-2mm},
            varwidth boxed title,
            colback=white,
            colframe=black,
            colbacktitle=white,
            coltitle=black,
            fonttitle=\bfseries,
            boxed title style={boxrule=1pt},
            boxrule=1pt,
            arc=5pt,
            left=2mm, right=2mm, top=2.5mm, bottom=1.5mm
        ]
        \setlength{\parskip}{2pt}
        \setlength{\parindent}{0pt}
        \setlist[itemize]{topsep=0pt,itemsep=1pt,parsep=0pt,partopsep=0pt}
        \setlist[enumerate]{topsep=0pt,itemsep=1pt,parsep=0pt,partopsep=0pt}
        \setlength{\abovedisplayskip}{2pt}
        \setlength{\belowdisplayskip}{0pt}
        \setlength{\abovedisplayshortskip}{2pt}
        \setlength{\belowdisplayshortskip}{0pt}

        \textbf{Notation.} $\prec$ is the canonical order of proof relations $\pi_i$ induced by the public state-transition trace.
        $\Xi(i)$ stores the fixed structural domain $\Omega_2(i)$, the current $\Omega(i)$ and $M(i)$, the guarded family $\Sigma(i)$, and the projected token/value labels attached to commitments created by $\pi_i$.
        $R$ is the shared reserved-commitment set and $Q$ is a fair work queue.
        $\mathsf{Init}$, $\mathsf{Refresh}$, $\mathsf{Lock}$, and $\mathsf{Notify}$ are defined in \appref{app:combined_propagation}.

        \textbf{function} $\Pi_\mathrm{prop}\bigl((\pi_i,\Omega_0(i))_{i=1}^n\bigr)$
        \begin{algorithmic}
        \State $Q \leftarrow \varnothing$
        \State $R \leftarrow \varnothing$
        \ForAll{$\pi_i$ in order $\prec$}
            \State $\Xi(i) \leftarrow \mathsf{Init}(i,\Omega_0(i))$
            \State $\mathsf{Lock}(i)$
            \State $Q \leftarrow Q \cup \mathsf{Notify}(i)$
        \EndFor
        \While{$Q\neq\varnothing$}
            \State $i \leftarrow \mathrm{pop}(Q)$
            \State $\mathsf{Refresh}(i)$
            \State $\mathsf{Lock}(i)$
            \State $Q \leftarrow Q \cup \mathsf{Notify}(i)$
        \EndWhile
        \State \Return $\{\Omega(i)\}_{i=1}^n,\{\Sigma(i)\}_{i=1}^n$ extracted from $\{\Xi(i)\}_{i=1}^n$
        \end{algorithmic}
        \textbf{end function}

    \end{tcolorbox}
    \caption{Cross-proof propagation to a common fixpoint.}
    \label{fig:alg_combinations}
\end{figure}

\point{Initialization}
As shown in \autoref{fig:alg_combinations}, $\mathsf{Init}(i,\Omega_0(i))$ constructs the state $\Xi(i)$ defined above. It fixes $\Omega_2(i)$ using Strategies~1 and~2, then evaluates the token--value tail under the reservations and projected labels available when $\pi_i$ is first reached in order $\prec$. As required by \autoref{ssec:s4}, unresolved hidden assignments do not by themselves remove candidates from the initialized guarded slices.

\point{Refresh}
$\mathsf{Refresh}(i)$ is used only after a notified update. It keeps $\Omega_2(i)$ and all immutable public parameters fixed, and recomputes the semantic tail under the current $R$ and current projected labels. A candidate may disappear because it is now reserved by another proof, because its token label no longer intersects a surviving branch, or because its value interval no longer supports the public balance relation. No other removal rule is available.

If refresh changes $\Theta(i)$, a guarded slice, $M(i)$, or an output label owned by $\pi_i$, the change is recorded before affected proofs are enqueued. If refresh makes every branch infeasible, the analysis terminates rather than selecting an interpretation unsupported by the public constraints.

\point{Lock}
$\mathsf{Lock}(i)$ adds every newly mandatory commitment in $M(i)$ to $R$. The definition of $M(i)$ already requires the commitment to appear in every surviving explanation, so locking it introduces no choice among feasible hidden assignments. If $|M(i)|=m_i$, all inputs of $\pi_i$ are uniquely determined and remain fixed thereafter. The simpler condition $|\Omega(i)|=m_i$ is sufficient for this case, but it is not the definition of mandatory status: commitments may be mandatory even when additional optional candidates remain.

Reservations are monotone. Once a commitment has been proved mandatory for one relation, later shrinkage cannot make it available to another without contradicting the first relation's surviving explanation set.

\point{Notification}
$\mathsf{Notify}(i)$ identifies proofs $\pi_j$ allowed by the instantiation's notification relation. It enqueues $\pi_j$ if its fixed structural domain contains a newly reserved commitment or if it contains a proof-created commitment whose projected label was narrowed by the current state of $\pi_i$. The latter case is necessarily forward because a created commitment can enter only a later authenticated state. Proofs that satisfy neither condition are not revisited.

\point{Monotonicity, Fixpoint, and Termination}
Propagation operates on a finite product state. The reserved set $R$ only grows; $\Omega(i)$ and its guarded slices only shrink; $M(i)$ only gains newly forced commitments; and $\Theta(i)$ and projected labels only narrow. No action can restore a candidate or interpretation already ruled out.

The updates are therefore monotone and contracting. Fair processing reaches the greatest common fixpoint below the initial states under the declared notification relation: queue order may change the number of revisits, but an effective contraction always re-enqueues every affected proof. Termination follows because each effective update strictly changes a finite component. When $Q$ is empty, the algorithm returns the final $\Omega(i)$ and $\Sigma(i)$ consumed by closure lifting in \autoref{ssec:closure_lifting}.

\begin{table}[t]
  \caption{Hinkal pool inventory; shading marks selected pools.}
  \label{tab:hinkal_deployments}
  \centering
  \scriptsize
  \setlength{\tabcolsep}{2.2pt}
  \renewcommand{\arraystretch}{0.95}
  \begin{adjustbox}{width=\linewidth}
    \begin{tabular}{@{}lcllrr@{}}
      \toprule
      Chain                                                         & Order                 & Contract / Program                                                                                                                   & Deployment Date (UTC)                                                                                                    & Tree Height            & Commitments                 \\
      \midrule
      \multirow{6}{*}{Ethereum}
                                                                    & $D1^{\dagger}$        & \href{https://etherscan.io/address/0x59a5c423bb7fd053d0356e6b46c2fbd3cda60c75}{\texttt{0x59a5...0c75}}                               & \href{https://etherscan.io/address/0x59a5c423bb7fd053d0356e6b46c2fbd3cda60c75}{2023-09-29}                               & 25                     & 23                          \\
                                                                    & D2                    & \href{https://etherscan.io/address/0x2ea81946ff675d5eb88192144ffc1418fa442e28}{\texttt{0x2ea8...2e28}}                               & \href{https://etherscan.io/address/0x2ea81946ff675d5eb88192144ffc1418fa442e28}{2023-11-10}                               & 25                     & 1{,}911                     \\
                                                                    & D3                    & \href{https://etherscan.io/address/0x13b65fa5375af51a12762499602be7b92afb8cd6}{\texttt{0x13b6...8cd6}}                               & \href{https://etherscan.io/address/0x13b65fa5375af51a12762499602be7b92afb8cd6}{2025-05-23}                               & 200                    & 30                          \\
                                                                    & \cellcolor{gray!15}D4 & \cellcolor{gray!15}\href{https://etherscan.io/address/0x25e5e82f5702a27c3466fe68f14abdbbadfca826}{\texttt{0x25e5...a826}}            & \cellcolor{gray!15}\href{https://etherscan.io/address/0x25e5e82f5702a27c3466fe68f14abdbbadfca826}{2025-06-27}            & \cellcolor{gray!15}200 & \cellcolor{gray!15}5{,}997  \\
                                                                    & $D5^{\dagger}$        & \href{https://etherscan.io/address/0xed03f6e7929ec5a81196634b9955a6838847ea1c}{\texttt{0xed03...ea1c}}                               & \href{https://etherscan.io/address/0xed03f6e7929ec5a81196634b9955a6838847ea1c}{2026-07-15}                               & 200                    & 7                           \\
                                                                    & D6                    & \href{https://etherscan.io/address/0x7cb60446d7635c68edf1c568cac74a1f98c1cfa4}{\texttt{0x7cb6...cfa4}}                               & \href{https://etherscan.io/address/0x7cb60446d7635c68edf1c568cac74a1f98c1cfa4}{2026-07-16}                               & 200                    & 159                         \\
      \midrule
      \multirow{5}{*}{Arbitrum}
                                                                    & $D1^{\dagger}$        & \href{https://arbiscan.io/address/0x8498a92fd1d6b32b2f388675904658909ecb456c}{\texttt{0x8498...456c}}                                & \href{https://arbiscan.io/address/0x8498a92fd1d6b32b2f388675904658909ecb456c}{2023-09-26}                                & 25                     & 140                         \\
                                                                    & D2                    & \href{https://arbiscan.io/address/0x41658b0daf59bb2fbb2d9a5249207011d2b364de}{\texttt{0x4165...64de}}                                & \href{https://arbiscan.io/address/0x41658b0daf59bb2fbb2d9a5249207011d2b364de}{2023-11-08}                                & 25                     & 20{,}835                    \\
                                                                    & D3                    & \href{https://arbiscan.io/address/0x13b65fa5375af51a12762499602be7b92afb8cd6}{\texttt{0x13b6...8cd6}}                                & \href{https://arbiscan.io/address/0x13b65fa5375af51a12762499602be7b92afb8cd6}{2025-05-23}                                & 200                    & 565                         \\
                                                                    & \cellcolor{gray!15}D4 & \cellcolor{gray!15}\href{https://arbiscan.io/address/0x25e5e82f5702a27c3466fe68f14abdbbadfca826}{\texttt{0x25e5...a826}}             & \cellcolor{gray!15}\href{https://arbiscan.io/address/0x25e5e82f5702a27c3466fe68f14abdbbadfca826}{2025-06-27}             & \cellcolor{gray!15}200 & \cellcolor{gray!15}8{,}579  \\
                                                                    & D5                    & \href{https://arbiscan.io/address/0x7cb60446d7635c68edf1c568cac74a1f98c1cfa4}{\texttt{0x7cb6...cfa4}}                                & \href{https://arbiscan.io/address/0x7cb60446d7635c68edf1c568cac74a1f98c1cfa4}{2026-07-16}                                & 200                    & 190                         \\
      \midrule
      \multirow{7}{*}{Polygon}
                                                                    & $D1^{\dagger}$        & \href{https://polygonscan.com/address/0xd7b7c69800f6e012f91c1b75b591991078a134d6}{\texttt{0xd7b7...34d6}}                            & \href{https://polygonscan.com/address/0xd7b7c69800f6e012f91c1b75b591991078a134d6}{2023-09-18}                            & 25                     & 3                           \\
                                                                    & $D2^{\dagger}$        & \href{https://polygonscan.com/address/0x26095cab8072258753f53a38a887815acd9364fb}{\texttt{0x2609...64fb}}                            & \href{https://polygonscan.com/address/0x26095cab8072258753f53a38a887815acd9364fb}{2023-09-22}                            & 25                     & 168                         \\
                                                                    & D3                    & \href{https://polygonscan.com/address/0xeeeeb52e36c78b153caab2761c369a50b066cdd5}{\texttt{0xeeee...cdd5}}                            & \href{https://polygonscan.com/address/0xeeeeb52e36c78b153caab2761c369a50b066cdd5}{2023-11-07}                            & 25                     & 15{,}956                    \\
                                                                    & D4                    & \href{https://polygonscan.com/address/0x13b65fa5375af51a12762499602be7b92afb8cd6}{\texttt{0x13b6...8cd6}}                            & \href{https://polygonscan.com/address/0x13b65fa5375af51a12762499602be7b92afb8cd6}{2025-05-23}                            & 200                    & 506                         \\
                                                                    & \cellcolor{gray!15}D5 & \cellcolor{gray!15}\href{https://polygonscan.com/address/0x25e5e82f5702a27c3466fe68f14abdbbadfca826}{\texttt{0x25e5...a826}}         & \cellcolor{gray!15}\href{https://polygonscan.com/address/0x25e5e82f5702a27c3466fe68f14abdbbadfca826}{2025-06-27}         & \cellcolor{gray!15}200 & \cellcolor{gray!15}13{,}692 \\
                                                                    & $D6^{\dagger}$        & \href{https://polygonscan.com/address/0xed03f6e7929ec5a81196634b9955a6838847ea1c}{\texttt{0xed03...ea1c}}                            & \href{https://polygonscan.com/address/0xed03f6e7929ec5a81196634b9955a6838847ea1c}{2026-07-15}                            & 200                    & 124                         \\
                                                                    & D7                    & \href{https://polygonscan.com/address/0x7cb60446d7635c68edf1c568cac74a1f98c1cfa4}{\texttt{0x7cb6...cfa4}}                            & \href{https://polygonscan.com/address/0x7cb60446d7635c68edf1c568cac74a1f98c1cfa4}{2026-07-16}                            & 200                    & 624                         \\
      \midrule
      \multirow{4}{*}{BNB Chain}
                                                                    & D1                    & \href{https://bscscan.com/address/0x0036e884cab4f427193839788edebb4b92b9a069}{\texttt{0x0036...a069}}                                & \href{https://bscscan.com/address/0x0036e884cab4f427193839788edebb4b92b9a069}{2023-11-08}                                & 25                     & 5{,}512                     \\
                                                                    & D2                    & \href{https://bscscan.com/address/0x13b65fa5375af51a12762499602be7b92afb8cd6}{\texttt{0x13b6...8cd6}}                                & \href{https://bscscan.com/address/0x13b65fa5375af51a12762499602be7b92afb8cd6}{2025-05-23}                                & 200                    & 0                           \\
                                                                    & D3                    & \href{https://bscscan.com/address/0x25e5e82f5702a27c3466fe68f14abdbbadfca826}{\texttt{0x25e5...a826}}                                & \href{https://bscscan.com/address/0x25e5e82f5702a27c3466fe68f14abdbbadfca826}{2025-06-27}                                & 200                    & 0                           \\
                                                                    & $D4^{\dagger}$        & \href{https://bscscan.com/address/0x7cb60446d7635c68edf1c568cac74a1f98c1cfa4}{\texttt{0x7cb6...cfa4}}                                & \href{https://bscscan.com/address/0x7cb60446d7635c68edf1c568cac74a1f98c1cfa4}{2026-07-29}                                & 200                    & 37                          \\
      \midrule
      \multirow{4}{*}{Base}
                                                                    & D1                    & \href{https://basescan.org/address/0x41658b0daf59bb2fbb2d9a5249207011d2b364de}{\texttt{0x4165...64de}}                               & \href{https://basescan.org/address/0x41658b0daf59bb2fbb2d9a5249207011d2b364de}{2024-03-22}                               & 25                     & 17{,}385                    \\
                                                                    & D2                    & \href{https://basescan.org/address/0x13b65fa5375af51a12762499602be7b92afb8cd6}{\texttt{0x13b6...8cd6}}                               & \href{https://basescan.org/address/0x13b65fa5375af51a12762499602be7b92afb8cd6}{2025-05-23}                               & 200                    & 611                         \\
                                                                    & \cellcolor{gray!15}D3 & \cellcolor{gray!15}\href{https://basescan.org/address/0x25e5e82f5702a27c3466fe68f14abdbbadfca826}{\texttt{0x25e5...a826}}            & \cellcolor{gray!15}\href{https://basescan.org/address/0x25e5e82f5702a27c3466fe68f14abdbbadfca826}{2025-06-27}            & \cellcolor{gray!15}200 & \cellcolor{gray!15}24{,}453 \\
                                                                    & D4                    & \href{https://basescan.org/address/0x7cb60446d7635c68edf1c568cac74a1f98c1cfa4}{\texttt{0x7cb6...cfa4}}                               & \href{https://basescan.org/address/0x7cb60446d7635c68edf1c568cac74a1f98c1cfa4}{2026-07-16}                               & 200                    & 767                         \\
      \midrule
      \multirow{5}{*}{Optimism}
                                                                    & $D1^{\dagger}$        & \href{https://optimistic.etherscan.io/address/0x436926c222bee240aa75b31443949d58c959d853}{\texttt{0x4369...d853}}                    & \href{https://optimistic.etherscan.io/address/0x436926c222bee240aa75b31443949d58c959d853}{2023-10-05}                    & 25                     & 24                          \\
                                                                    & D2                    & \href{https://optimistic.etherscan.io/address/0x41658b0daf59bb2fbb2d9a5249207011d2b364de}{\texttt{0x4165...64de}}                    & \href{https://optimistic.etherscan.io/address/0x41658b0daf59bb2fbb2d9a5249207011d2b364de}{2023-11-08}                    & 25                     & 11{,}555                    \\
                                                                    & D3                    & \href{https://optimistic.etherscan.io/address/0x13b65fa5375af51a12762499602be7b92afb8cd6}{\texttt{0x13b6...8cd6}}                    & \href{https://optimistic.etherscan.io/address/0x13b65fa5375af51a12762499602be7b92afb8cd6}{2025-05-23}                    & 200                    & 2{,}477                     \\
                                                                    & \cellcolor{gray!15}D4 & \cellcolor{gray!15}\href{https://optimistic.etherscan.io/address/0x25e5e82f5702a27c3466fe68f14abdbbadfca826}{\texttt{0x25e5...a826}} & \cellcolor{gray!15}\href{https://optimistic.etherscan.io/address/0x25e5e82f5702a27c3466fe68f14abdbbadfca826}{2025-06-27} & \cellcolor{gray!15}200 & \cellcolor{gray!15}25{,}132 \\
                                                                    & D5                    & \href{https://optimistic.etherscan.io/address/0x969ef3a23f8cf732655c70835f86c2d149748188}{\texttt{0x969e...8188}}                    & \href{https://optimistic.etherscan.io/address/0x969ef3a23f8cf732655c70835f86c2d149748188}{2026-07-23}                    & 200                    & 4                           \\
      \midrule
      \multirow{4}{*}{Avalanche}
                                                                    & $D1^{\dagger}$        & \href{https://snowtrace.io/address/0x390f9ff1c00021d77a30ee1d59ee3e04f9b0322f}{\texttt{0x390f...322f}}                               & \href{https://snowtrace.io/address/0x390f9ff1c00021d77a30ee1d59ee3e04f9b0322f}{2023-09-26}                               & 25                     & 116                         \\
                                                                    & D2                    & \href{https://snowtrace.io/address/0x41658b0daf59bb2fbb2d9a5249207011d2b364de}{\texttt{0x4165...64de}}                               & \href{https://snowtrace.io/address/0x41658b0daf59bb2fbb2d9a5249207011d2b364de}{2023-11-09}                               & 25                     & 2{,}284                     \\
                                                                    & D3                    & \href{https://snowtrace.io/address/0x13b65fa5375af51a12762499602be7b92afb8cd6}{\texttt{0x13b6...8cd6}}                               & \href{https://snowtrace.io/address/0x13b65fa5375af51a12762499602be7b92afb8cd6}{2025-05-23}                               & 200                    & 197                         \\
                                                                    & D4                    & \href{https://snowtrace.io/address/0x25e5e82f5702a27c3466fe68f14abdbbadfca826}{\texttt{0x25e5...a826}}                               & \href{https://snowtrace.io/address/0x25e5e82f5702a27c3466fe68f14abdbbadfca826}{2025-06-27}                               & 200                    & 280                         \\
      \midrule
      Tempo
                                                                    & D1                    & \href{https://explore.tempo.xyz/address/0x631c98336d4684333946262a9e3e942ef3555a5e}{\texttt{0x631c...5a5e}}                          & \href{https://explore.tempo.xyz/address/0x631c98336d4684333946262a9e3e942ef3555a5e}{2026-06-02}                          & 200                    & 455                         \\
      \midrule
      Tron
                                                                    & D1                    & \href{https://tronscan.org/\#/contract/TKFUxULu53pSfDkSZwF85PFuKBw1K9axaw}{\texttt{TKFU...9axaw}}                                    & \href{https://tronscan.org/\#/contract/TKFUxULu53pSfDkSZwF85PFuKBw1K9axaw}{2026-03-16}                                   & 200                    & 61                          \\
      \midrule
      \multirow{2}{*}{Solana}
                                                                    & D1                    & \href{https://explorer.solana.com/address/BD7ycVagzvCbKU9xb8DkzZJHViwSLy1CLkGG6NPc95ER}{\texttt{BD7y...95ER}}                        & \href{https://explorer.solana.com/address/BD7ycVagzvCbKU9xb8DkzZJHViwSLy1CLkGG6NPc95ER}{2025-12-30}$^{\ddagger}$         & 200                    & 48                          \\
                                                                    & D2                    & \href{https://explorer.solana.com/address/J4SsjA1Zqf2tZfBJjYrEKXocM9NdP2xHNfAQLM7McG5H}{\texttt{J4Ss...cG5H}}                        & \href{https://explorer.solana.com/address/J4SsjA1Zqf2tZfBJjYrEKXocM9NdP2xHNfAQLM7McG5H}{2026-03-12}$^{\ddagger}$         & 200                    & 6{,}352                     \\
      \midrule
      \multicolumn{5}{@{}l}{Total: 39 deployed pools (37 nonempty)} & 167{,}259                                                                                                                                                                                                                                                                                                                                      \\
      \bottomrule
    \end{tabular}
  \end{adjustbox}
  \vspace{1pt}
  \parbox{\linewidth}{\scriptsize
    \setlength{\parindent}{0pt}
    $^{\dagger}$ Pool identified and verified directly on-chain, supplementing the recovered public deployment records for that period.\par
    $^{\ddagger}$ Solana deployment date inferred from the earliest visible program signature as the closest observable on-chain timestamp.}
\end{table}

\subsection{Hinkal Deployments and Selection}\label{app:hinkal_deployments}

Each independently deployed Hinkal pool maintains its own commitment tree and therefore forms a separate anonymity domain. \autoref{tab:hinkal_deployments} inventories these states and the selection used in \autoref{sec:empirical}. Labels $D1,D2,\ldots$ give deployment order within a chain; they are not cross-chain version identifiers.

At the inclusive cutoff (July~30,~2026, 23:59:59 UTC), 39 pool states are deployed across ten chains. The 37 nonempty pools contain 167{,}259 commitments, but activity is fragmented: 23 contain fewer than 1{,}000 commitments. A second Tempo deployment on July~31,~2026, at 08:26:34 UTC falls outside the window.

We combine public deployment records with direct chain histories and verify every listed pool against its commitment-tree state. For \acs{EVM}-compatible pools, the commitment count is the cutoff state index minus the initial leaf-index offset, cross-checked against commitment-insertion events. ``Tree Height'' is the deployed tree's level parameter, not its observed leaf count. Tron events and Solana signatures use the same UTC cutoff. All explorer links were accessed August~15,~2026.

The five shaded rows are the most active pools in the coordinated \acs{EVM}-compatible generation deployed on June~27,~2025. That generation contains 78{,}133 commitments across its \acs{EVM}-compatible pools at the cutoff---the largest aggregate among Hinkal's coordinated cross-chain generations---while its BNB Chain and Avalanche pools contain zero and 280, respectively. Fixing one generation keeps the implementation comparable across chains; no pool from another generation is merged into an evaluated anonymity domain.

\subsection{Protocol Instantiations}
\label{app:protocol-instantiations}

\begin{figure}[t]
    \centering
    \footnotesize
    \begin{tcolorbox}[
            enhanced,
            title=Railgun V2 JoinSplit proof,
            attach boxed title to top left={xshift=2mm,yshift=-2mm},
            varwidth boxed title,
            colback=white,
            colframe=black,
            colbacktitle=white,
            coltitle=black,
            fonttitle=\bfseries,
            boxed title style={boxrule=1pt},
            boxrule=1pt,
            arc=5pt,
            left=2mm, right=2mm, top=2mm, bottom=1.2mm
        ]
        \setlength{\parskip}{1pt}
        \setlength{\parindent}{0pt}
        \setlist[itemize]{topsep=0pt,itemsep=1pt,parsep=0pt,partopsep=0pt}
        \setlength{\abovedisplayskip}{1pt}
        \setlength{\belowdisplayskip}{0pt}
        \setlength{\abovedisplayshortskip}{1pt}
        \setlength{\belowdisplayshortskip}{0pt}

        Public input: a Merkle root $R$, a bound-parameter hash $h_{\mathrm{bp}}$, input nullifiers $(n_i)_{i=1}^m$, and circuit output commitments $(c_j^{\mathrm{out}})_{j=1}^q$.

        Witness: a common token identifier $\tau$, a spending public key $PK$, a signature $\sigma$, a nullifying key $\mathsf{nk}$, input-note tuples $(\rho_i,v_i,\ell_i,\mathsf{path}_i)_{i=1}^m$, and output-note tuples $(\mathsf{npk}^{\mathrm{out}}_j,v_j^{\mathrm{out}})_{j=1}^q$, where $\ell_i$ is the input leaf index. $\mathsf{Poseidon}$ denotes the appropriate-arity hash.

        Compute the signed public message $h$ as the Poseidon hash of $R$, $h_{\mathrm{bp}}$, all input nullifiers, and all output commitments.

        Check authorization and binding:
        \begin{itemize}
            \item verify the signature $\sigma$ on $h$ under $PK$;
            \item because $h$ includes $h_{\mathrm{bp}}$, the proof is bound to the transaction parameters as well as its nullifiers and output commitments;
            \item compute the master public key $\mathsf{mpk}=\mathsf{Poseidon}(PK_x,PK_y,\mathsf{nk})$;
            \item use the same $\tau$ in every input and output commitment, so all consumed and created notes share one asset type, as in \autoref{ssec:proof}.
        \end{itemize}

        For each input note $i\in\{1,\dots,m\}$:
        \begin{itemize}
            \item derive the nullifier $n_i=\mathsf{Poseidon}(\mathsf{nk},\ell_i)$;
            \item derive the input note public key $\mathsf{npk}^{\mathrm{in}}_i=\mathsf{Poseidon}(\mathsf{mpk},\rho_i)$;
            \item reconstruct $c_i=\mathsf{Poseidon}(\mathsf{npk}^{\mathrm{in}}_i,\tau,v_i)$;
            \item verify with $\mathsf{path}_i$ that $c_i$ belongs to the tree rooted at $R$.
        \end{itemize}

        For each output note $j\in\{1,\dots,q\}$:
        \begin{itemize}
            \item require $v_j^{\mathrm{out}}$ to fit into the prescribed bit range;
            \item compute $\widehat{c}_j^{\mathrm{out}}=\mathsf{Poseidon}(\mathsf{npk}^{\mathrm{out}}_j,\tau,v_j^{\mathrm{out}})$;
            \item require $\widehat{c}_j^{\mathrm{out}}=c_j^{\mathrm{out}}$.
        \end{itemize}

        Require value conservation, $\sum_{i=1}^{m}v_i=\sum_{j=1}^{q}v_j^{\mathrm{out}}$.

        If a plaintext output is present, the contract treats the final circuit output as $O_i$ and omits it from Merkle insertion; hence $k_i=q-1$, and otherwise $k_i=q$, as defined in \autoref{ssec:proof}.

        Accept if all checks above hold.
    \end{tcolorbox}
    \caption{Railgun V2 JoinSplit proof.}
    \label{fig:alg_railgun_v2_joinsplit}
\end{figure}

\begin{table*}[t]
    \centering
    \footnotesize
    \setlength{\tabcolsep}{5pt}
    \renewcommand{\arraystretch}{1.16}
    \caption{Instantiation of the system model and analysis in Railgun and Hinkal.}
    \label{tab:protocol_instantiation_mapping}
    \begin{tabular}{@{}p{0.18\textwidth}p{0.37\textwidth}p{0.37\textwidth}@{}}
        \toprule
        \textbf{Model element}
          & \textbf{Railgun \href{https://app.dedaub.com/ethereum/address/0xbcfa4de73afb071c9ff18a20a22f818e657c541a/source}{V1}/\href{https://github.com/Railgun-Privacy/contract}{V2}}
          & \textbf{\href{https://etherscan.io/address/0x25e5e82f5702a27c3466fe68f14abdbbadfca826\#code}{Hinkal pools deployed June 27, 2025}} \\
        \midrule
        {\raggedright\textbf{Authenticated pool state}\par}
          & Each deployment maintains a sequence of fixed-capacity Merkle trees; $\pi_i$ identifies one tree and one authenticated historical root.
          & Each evaluated pool maintains one independent commitment tree; $\pi_i$ identifies a historical state through the call's authenticated root, without a tree-number partition. \\
        {\raggedright\textbf{Shield and\newline shielding address}\par}
          & Shield transfers assets into $\mathcal{P}$, and its public trace directly identifies the shielding address.
          & Shield may use the proofless entry point or be part of the unified call. The protocol-executed transfer into $\mathcal{P}$ identifies the shielding address, which may differ from the submitter. \\
        {\raggedright\textbf{Proof granularity ($\pi_i$)}\par}
          & One JoinSplit proof instantiates one same-token input--output relation.
          & One cryptographic proof may jointly authenticate several same-token relations, each represented by a separate $\pi_i$. \\
        {\raggedright\textbf{Token and\newline Value constraints}\par}
          & A JoinSplit uses one common token; public Shield/Unshield data, commitment-creation histories, and the balance relation determine Token and Value feasibility.
          & The exact public token address and balance change of each same-token relation determine Token and Value feasibility. \\
        {\raggedright\textbf{Plaintext output $O_i$}\par}
          & A V1 withdrawal preimage or V2 Unshield output instantiates $O_i$ and records the token and value released from $\mathcal{P}$.
          & The public outflow to the unshielding address instantiates $O_i$; a protocol-distinguished relayer fee does not. \\
        {\raggedright\textbf{Commitment provenance for closure}\par}
          & A commitment is introduced by Shield or a JoinSplit; closure follows a proof-created commitment through its creating proof under the same token guard.
          & A commitment may have a Shield source, private predecessors, or both. Closure changes token guards only when the public execution trace links the corresponding input and output tokens. \\
        {\raggedright\textbf{Transaction boundary}\par}
          & A spend transaction may aggregate JoinSplit proofs, all checked before any private commitments are inserted.
          & A unified call may jointly authenticate several relations; verification and public execution complete before its new commitments are inserted. \\
        \bottomrule
    \end{tabular}
\end{table*}

This appendix maps the model of \autoref{sec:system} and the analysis of \autoref{sec:deanonymization} to the deployed protocols evaluated in \autoref{sec:empirical}. \autoref{tab:protocol_instantiation_mapping} summarizes the correspondence. Railgun V2 publishes contract and circuit source; for Railgun V1 and the evaluated Hinkal pools, the analysis uses contract-visible semantics and public proof inputs without inferring witness constraints absent from the available source. All implementation links below were accessed August~15,~2026.

\point{Railgun V1 and V2}
We identify the production proxies from Railgun's \href{https://docs.railgun.org/wiki/learn/helpful-links}{deployment registry} and inspect the \href{https://app.dedaub.com/ethereum/address/0xbcfa4de73afb071c9ff18a20a22f818e657c541a/source}{deployed V1 contract} and the public V2 \href{https://github.com/Railgun-Privacy/contract}{contract} and \href{https://github.com/Railgun-Privacy/circuits-v2}{circuit} repositories. On the three deployments spanning both versions, V2 continues through the same production proxy and preserves the existing commitment-tree state, so we reconstruct one continuous history rather than separate version-specific pools. \autoref{fig:alg_railgun_v2_joinsplit} summarizes the model-relevant V2 circuit constraints.

At the contract-visible boundary, V1 Shield validates public note preimages, transfers their assets into the pool, and records the derived commitments. A V1 spend requires a previously seen root and unused nullifiers before proof verification. If a public withdrawal preimage is present, its hash must equal the final output commitment; the contract releases that value, excludes the final output from Merkle insertion, and inserts the remaining private outputs.

Both Shield paths expose the created note preimages while inserting only their commitment hashes. Each Railgun JoinSplit proof instantiates one proof relation $\pi_i$: it uses one tree and authenticated root, consumes notes of one token, and may create private commitments plus one plaintext output $O_i$. V1 represents $O_i$ by a public withdrawal preimage, whereas V2 exposes it through the Unshield path. Its commitment is the final circuit output but is excluded from Merkle insertion, so $k_i$ in \autoref{ssec:proof} counts only private outputs. All proofs in a spend transaction are checked before any of their private outputs are inserted, matching \autoref{ssec:transaction}.

The reconstruction uses the public events that record Shield commitments, nullifiers, proof-created commitments, and Unshield outputs. V2's JoinSplit circuit additionally enforces the common-token relation used by \autoref{ssec:s3}. Administrative events are excluded because they do not alter note creation, consumption, or public release.

The model is unchanged across versions, but reconstruction accounts for V2's more specific events and additional token classes, and for the different exposure of the public recipient. Because the V1 circuit is unavailable, its evidence boundary remains contract-visible; \appref{app:analysis_validation} reports the corresponding full-trace checks.

\point{Hinkal}
Each Hinkal contract selected in \appref{app:hinkal_deployments} maintains an independent pool and commitment tree. We inspect the verified \href{https://etherscan.io/address/0x25e5e82f5702a27c3466fe68f14abdbbadfca826\#code}{pool contract} and execution contracts active at \href{https://etherscan.io/address/0xccdd76c688cd3d8f22e602c3a215227daa313096\#code}{deployment} and the \href{https://etherscan.io/address/0xde88ff7ba99ac02109f4604bef2b48e01be27a4b\#code}{cutoff}. We found neither verified source for the contract that assembles the verifier's public inputs nor public circuit source for these pools. The mapping therefore uses only the observable proof-call boundary and infers no hidden circuit constraint. We treat this as an evidence boundary rather than inferred circuit semantics: no unavailable witness constraint is used to eliminate a candidate. Additional witness constraints could make the exact feasible set smaller than the reported set, but every reported elimination is justified by an observable proof-call input, a contract-enforced state transition, or a corresponding asset transfer. The Hinkal results are therefore conservative with respect to unavailable circuit semantics and do not constitute an audit of circuit completeness or verifier soundness.

The pool exposes a proofless Shield entry point (\texttt{prooflessDeposit}) and a unified proof-verified entry point (\texttt{transact}) that may combine private inputs and outputs with Shield, Unshield, or external execution. A unified call submits one cryptographic proof and publicly groups its nullifiers, created commitments, and balance changes by token. The model represents its input-bearing same-token relations as distinct proof relations $\pi_i$. This is an analytical projection of the public call, not a decomposition of its cryptographic proof: it preserves the granularity of \autoref{ssec:proof} without assuming separate or sequential verification, and all relations contributing to a public release are reunited at the transaction layer.

A public outflow may instead consume value that an earlier private transition placed in persistent authorization state. We model this as a same-token relation whose direct predecessor is that earlier transition, even when the current relation records no nullifier: closure follows $\mathsf{Pred}_{\pi}(i)$ and never treats the outflow as a new Shield source. The relation is Proof~\fcirc{2} without a private commitment and Proof~\fcirc{4} otherwise.

For these proof relations, Tree Number is inapplicable because the pool exposes no Railgun-style tree number; Proof Root uses the prefix covered by the authenticated historical root; Token uses the relation's exact public token address; and Value uses its public balance change and the contract-enforced balance equation. These are precisely the protocol-visible constraints of \autoref{ssec:note_source_pruning}.

The unified path also explains the provenance extension in \autoref{ssec:closure_lifting}: a created commitment may depend on both a Shield source and private predecessors, so $\mathsf{Entry}(c,t)$ and $\mathsf{Pred}(c,t)$ may both be nonempty. When public external execution relates different token branches, closure follows every permitted cross-token predecessor; otherwise it preserves the token guard. On Unshield, recipient value and the relayer fee are separate contract-governed outflows, so a fee-only branch is excluded from $O_i$ without treating the transaction submitter as a shielding or unshielding address.

Pool state persists across auxiliary-contract changes, which we replay and validate in ledger order (\appref{app:analysis_validation}).

\subsection{Reconstruction and Analysis Validation}
\label{app:analysis_validation}

\begin{figure}[t]
    \centering
    \begin{subfigure}{\linewidth}
        \begin{lstlisting}[style=artifactpatch,language=ArtifactRust,aboveskip=0pt,belowskip=0pt,framesep=3pt,numbersep=5pt,xleftmargin=0.9em,xrightmargin=0pt]
let proof_relation_id = proof_relation_by_nullifier
    .get(&nullifier_key).copied().ok_or_else(|| {
    format!("missing proof relation for nullifier {}", nullifier)
})?;
let proof_root = root_by_nullifier
    .get(&nullifier_key).cloned().ok_or_else(|| {
    format!("missing exact proof root for nullifier {}", nullifier)
})?;

// Later, while constructing the commitment-creator map:
match creator_by_commitment.get(&commitment_key) {
    None => {
        creator_by_commitment.insert(
            commitment_key.clone(), creating_proof.clone());
    }
    Some(existing) if existing == &creating_proof => {}
    Some(existing) => {
        return Err(format!(
            "conflicting creators for {}: {:?} vs {:?}",
            commitment, existing, creating_proof));
    }
}
\end{lstlisting}
        \caption{Exact nullifier/root coverage and creating-proof uniqueness.}
        \label{fig:code_graph_checks}
    \end{subfigure}

    \vspace{2mm}
    \begin{subfigure}{\linewidth}
        \begin{lstlisting}[style=artifactpatch,language=ArtifactRust,aboveskip=0pt,belowskip=0pt,framesep=3pt,numbersep=5pt,xleftmargin=0.9em,xrightmargin=0pt]
let feasible_tokens = feasible_tokens_by_tree
    .get(tree_id)
    .and_then(|per_tree| per_tree.get(leaf_index))
    .cloned()
    .ok_or_else(|| format!(
        "missing token labels for created output {}", leaf_index))?;
if feasible_tokens.is_empty() {
    return Err("empty token labels on created output".into());
}

let mut tokens_without_bounds = Vec::new();
for token in &feasible_tokens {
    if !projected_value_bounds.contains_key(token) {
        tokens_without_bounds.push(token.clone());
    }
}
if !tokens_without_bounds.is_empty() {
    return Err(format!(
        "created-output value bounds missing tokens: {:?}",
        tokens_without_bounds));
}
\end{lstlisting}
        \caption{Completeness of labels projected through created outputs.}
        \label{fig:code_propagation_checks}
    \end{subfigure}

    \caption{Representative fail-closed checks for Railgun.}
    \label{fig:code_consistency_checks}
\end{figure}

\begin{figure}[t]
    \centering
    \begin{subfigure}{\linewidth}
        \begin{lstlisting}[style=artifactpatch,language=ArtifactJavaScript,aboveskip=0pt,belowskip=0pt,framesep=3pt,numbersep=5pt,xleftmargin=0.9em,xrightmargin=0pt]
const candidates = [];
for (let leafIndex = 0; leafIndex < authenticatedLeafCount; leafIndex += 1) {
  const commitment = commitments[leafIndex];
  if (commitment.tokenAddress !== proofRelation.tokenAddress) continue;
  if (hasValueConstraint) {
    const label = valueLabelByLeaf.get(leafIndex);
    if (label && String(label.amountFieldValue) !== requiredInputValue)
      continue;
  }
  candidates.push(leafIndex);
}
const setDigest = digestCommitmentIndices(candidates);
requireCondition(candidates.length === storedPruningResult.size, "set size mismatch");
requireCondition(setDigest === storedPruningResult.digest, "set digest mismatch");
requireCondition(candidates.length > 0, "empty Commitment Set");

const priorSet = commitmentSetByProof.get(proofRelationId);
if (priorSet) requireCondition(
  priorSet.digest === setDigest &&
  compareArrays(priorSet.indices, candidates),
  "same-proof Commitment Sets differ");
\end{lstlisting}
        \caption{Independent reconstruction of a Hinkal Commitment Set.}
        \label{fig:code_domain_checks}
    \end{subfigure}

    \vspace{2mm}
    \begin{subfigure}{\linewidth}
        \begin{lstlisting}[style=artifactpatch,language=ArtifactJavaScript,aboveskip=0pt,belowskip=0pt,framesep=3pt,numbersep=5pt,xleftmargin=0.9em,xrightmargin=0pt]
function ledgerPosition(event) {
  return [event.blockNumber, event.transactionIndex,
          event.callIndex ?? -1, event.logIndex ?? -1].map(Number);
}
function precedes(left, right) {
  const a = ledgerPosition(left), b = ledgerPosition(right);
  for (let i = 0; i < a.length; i += 1)
    if (a[i] < b[i]) return true;
    else if (a[i] > b[i]) return false;
  return false;
}
function deriveMetrics(proof, sources) {
  for (const source of sources)
    requireCondition(precedes(source, proof), "non-prior source");
  const sourceTxs = new Set(sources.map((s) => s.transactionHash));
  const anonymitySet = new Set(sources.flatMap((s) => s.shieldingAddresses));
  return { sourceTxs, anonymitySet };
}
\end{lstlisting}
        \caption{Ledger-order validation and transaction-level address projection.}
        \label{fig:code_closure_checks}
    \end{subfigure}

    \caption{Representative fail-closed checks for Hinkal.}
    \label{fig:hinkal_code_consistency_checks}
\end{figure}

This appendix records the fail-closed checks at the boundaries that determine the reported Commitment and Anonymity Sets: any missing or inconsistent input terminates the analysis. \autoref{fig:code_consistency_checks} and \autoref{fig:hinkal_code_consistency_checks} show representative excerpts from the analysis and its independent validation. The excerpts preserve the executed checks while aligning identifiers with model terminology and omitting setup and diagnostics; every check runs over the complete analyzed datasets.

\point{Trace Reconstruction and Graph Integrity}
The analysis admits a proof relation only when every public nullifier has an exact decoded relation and authenticated root, and admits a proof-created commitment only when its creating proof is unique. \autoref{fig:code_graph_checks} shows both Railgun checks. Root lookup uses the reconstructed history of the referenced tree; a missing root or tree mismatch terminates the analysis before pruning. Hinkal additionally requires contiguous commitment insertion, one decoded call for each cryptographic proof, unique proof-relation identifiers, exact reconstruction of every root-covered prefix, and one creating proof for each proof-created commitment. These checks cover the datasets in \autoref{tab:cross_protocol_dataset_overview}: 497{,}034 commitments and 431{,}198 nullifiers for Railgun, and 77{,}853 commitments and 74{,}499 nullifiers for Hinkal.

\point{Propagation-State Consistency}
Projected token/value labels are shared state: when an earlier proof narrows, every later candidate created by it must observe the same update. The analysis therefore associates each projected output label with its creating proof, synchronizes it after a reservation or label contraction, and rejects an empty token set or a value map missing any surviving token guard. \autoref{fig:code_propagation_checks} shows the latter checks. This prevents an outdated or incomplete output label from surviving merely because the commitment was inserted before its creating proof reached the fixpoint. For Hinkal, closure independently reconstructs every local Value-stage Commitment Set from the root-covered prefix, exact public token address, and value-stage labels. As shown in \autoref{fig:code_domain_checks}, its size and a deterministic digest of its ordered commitment indices must match the stored local-pruning result. Validation then replays chronological mandatory-input reservations to reconstruct the final $\Omega(i)$. Nullifiers in the same proof relation must reconstruct identical sets, and an empty set terminates the analysis; neither stage trusts the closure output.

\point{Closure and Transaction-Level Lifting}
Every terminal Shield source reached by closure must precede the analyzed proof in canonical ledger order, including intra-transaction call and log order, as illustrated in \autoref{fig:code_closure_checks}. Railgun analogously checks each source against the reconstructed spend boundary and rejects unmapped or non-prior shield commitments. Address lifting deduplicates the reached shield transactions and shielding addresses; transactions with multiple contributing proofs use one common temporal baseline before union. Independently computed depth records must match all 169{,}840 Railgun and 16{,}516 Hinkal unshielding transactions.

\point{Protocol-Specific Evidence Boundaries}
Railgun V2 is checked against both its public contracts and JoinSplit circuit. Because the V1 circuit is unavailable, V1 validation is limited to contract-visible semantics but covers every decoded V1 call: 3{,}452 spend transactions containing 4{,}482 proof relations. Every rebuilt root predates its spend, every encoded proof and ciphertext array satisfies the V1 contract's format checks, all 5{,}483 nullifiers and 9{,}080 inserted commitments match their events, and all 2{,}343 withdrawal preimages hash to the final output commitment excluded from insertion.

Hinkal external execution can transfer value from private notes into persistent authorization state before a later transaction consumes it. Validation replays every such state change in ledger order, matches it to the decoded call and emitted event, and requires each debit to be supported by prior state. It also validates Shield provenance and fee-only branches against contract semantics and asset transfers. Where the public trace permits multiple predecessor transitions, closure retains all of them rather than selecting one. Any missing mapping, inconsistent state, incomplete metric match, or unsupported branch invalidates the analysis. All checks passed on the reported datasets.

\begin{figure*}[t]
    \centering
    \begin{subfigure}[t]{0.8\textwidth}
        \centering
        \includegraphics[width=\linewidth, trim=0 10 10 0, clip]{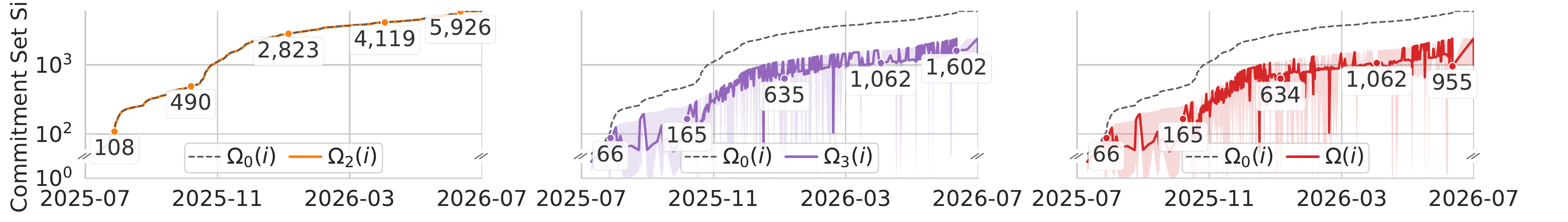}
        \caption{Ethereum}
    \end{subfigure}

    \vspace{0.4em}

    \begin{subfigure}[t]{0.8\textwidth}
        \centering
        \includegraphics[width=\linewidth, trim=0 10 10 0, clip]{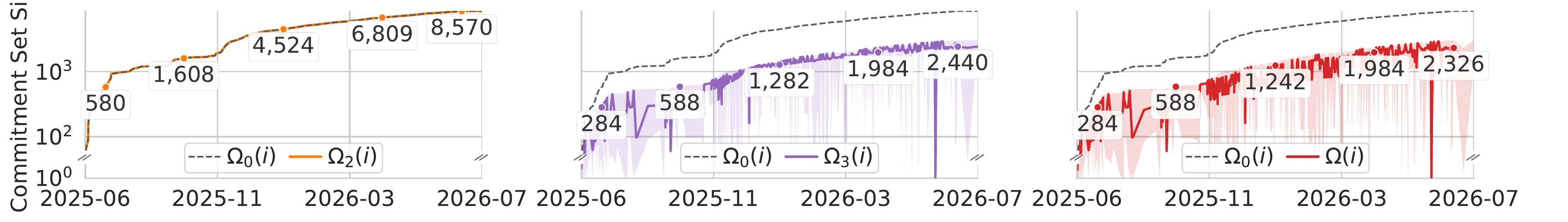}
        \caption{Arbitrum}
    \end{subfigure}

    \vspace{0.4em}

    \begin{subfigure}[t]{0.8\textwidth}
        \centering
        \includegraphics[width=\linewidth, trim=0 10 10 0, clip]{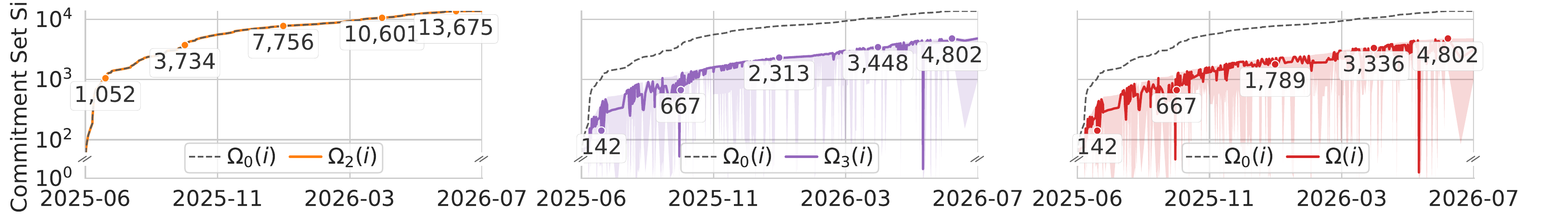}
        \caption{Polygon}
    \end{subfigure}

    \vspace{0.4em}

    \begin{subfigure}[t]{0.8\textwidth}
        \centering
        \includegraphics[width=\linewidth, trim=0 10 10 0, clip]{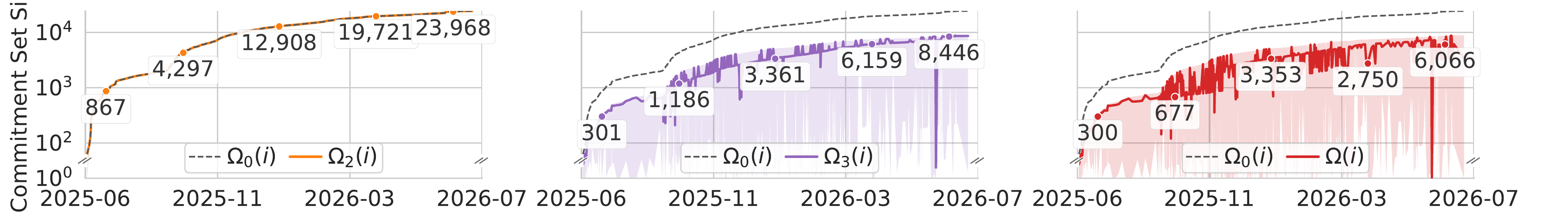}
        \caption{Base}
    \end{subfigure}

    \vspace{0.4em}

    \begin{subfigure}[t]{0.8\textwidth}
        \centering
        \includegraphics[width=\linewidth, trim=0 10 10 0, clip]{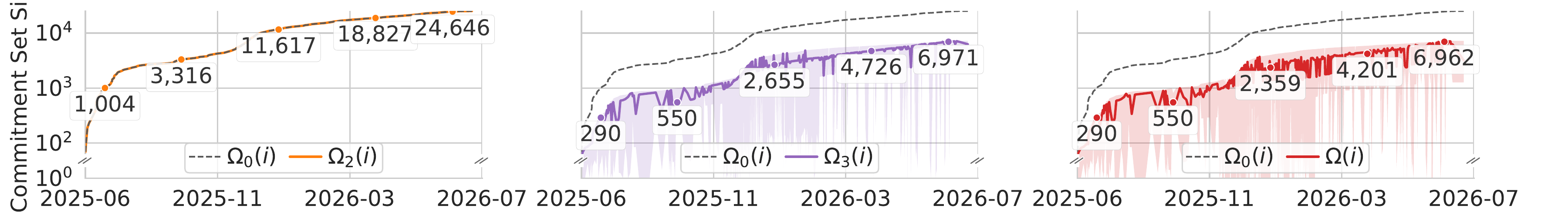}
        \caption{Optimism}
    \end{subfigure}

    \caption{Temporal evolution of Hinkal Commitment Set Size through July~30,~2026 (UTC), in 500 equal-count block-order bins per selected pool. Successive panels compare $\Omega_0(i)$ with $\Omega_2(i)$, $\Omega_3(i)$, and the propagated final $\Omega(i)$ after Value; Tree Number is inapplicable. Curves show P50; shading spans P0--P100; labels mark selected times.}
    \label{fig:hinkal_commitment_set_size_multichain}
\end{figure*}

\begin{figure*}[t]
    \centering
    \begin{subfigure}[t]{0.199\textwidth}
        \centering
        \includegraphics[width=\linewidth, trim=0 10 0 0, clip]{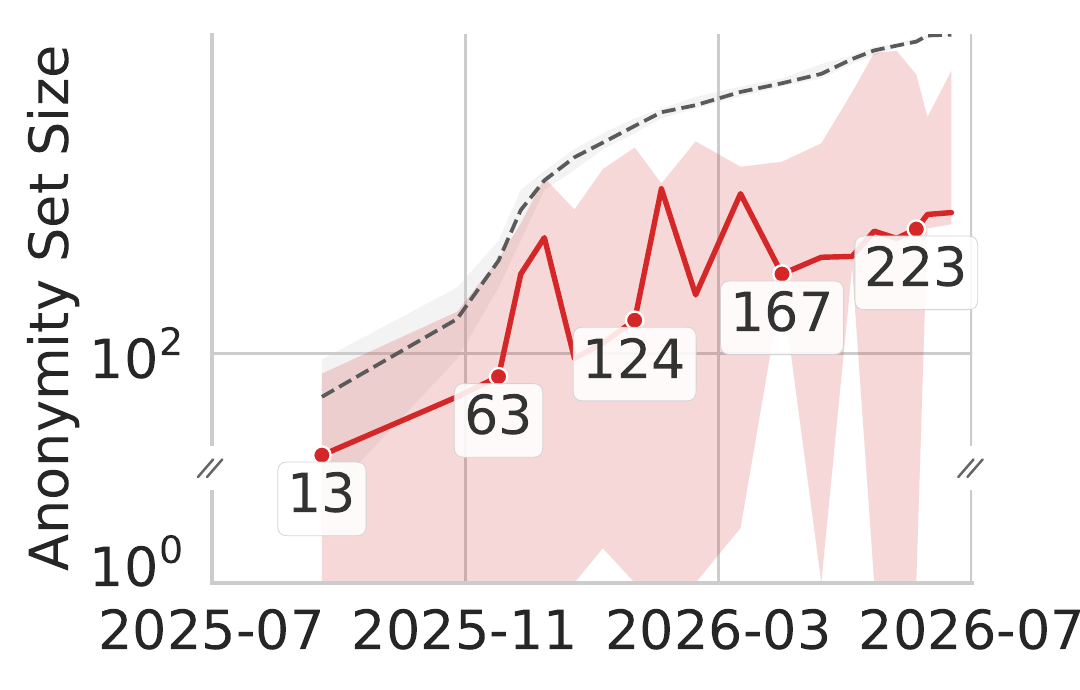}
        \caption{Ethereum}
    \end{subfigure}\hspace{0.001\textwidth}%
    \begin{subfigure}[t]{0.199\textwidth}
        \centering
        \includegraphics[width=\linewidth, trim=0 10 0 0, clip]{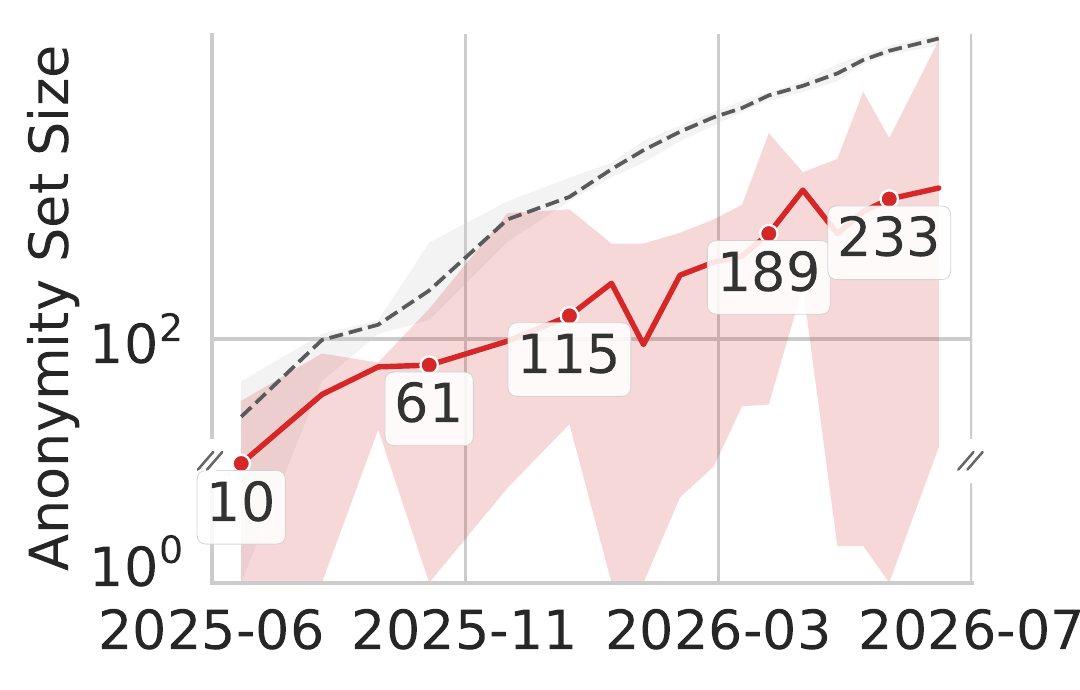}
        \caption{Arbitrum}
    \end{subfigure}\hspace{0.001\textwidth}%
    \begin{subfigure}[t]{0.199\textwidth}
        \centering
        \includegraphics[width=\linewidth, trim=0 10 0 0, clip]{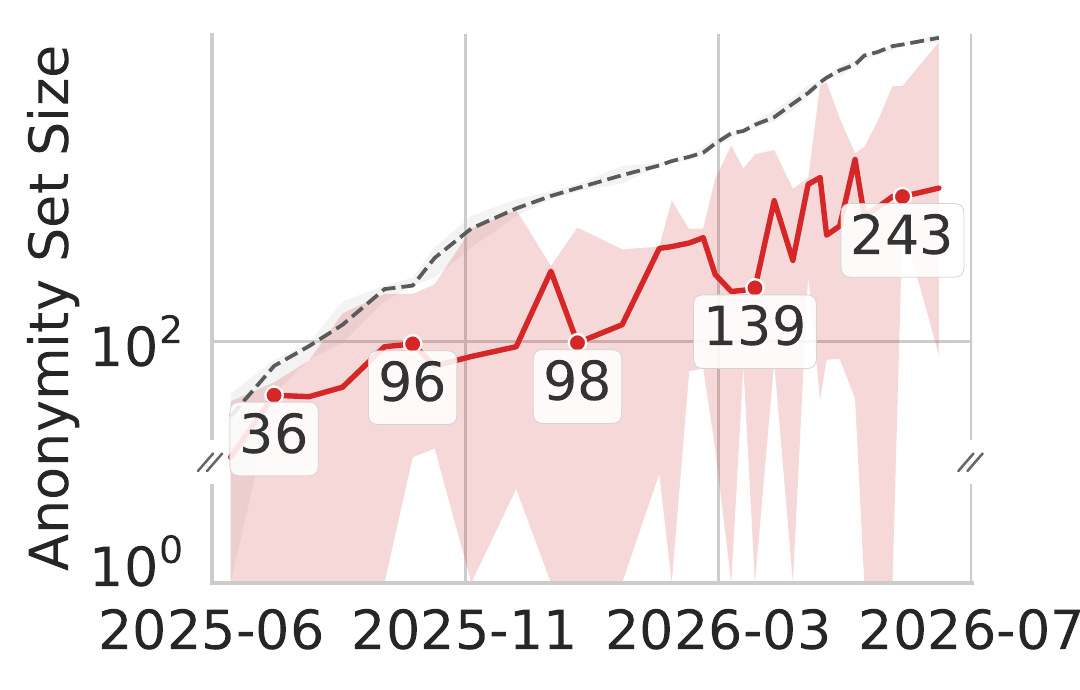}
        \caption{Polygon}
    \end{subfigure}\hspace{0.001\textwidth}%
    \begin{subfigure}[t]{0.199\textwidth}
        \centering
        \includegraphics[width=\linewidth, trim=0 10 0 0, clip]{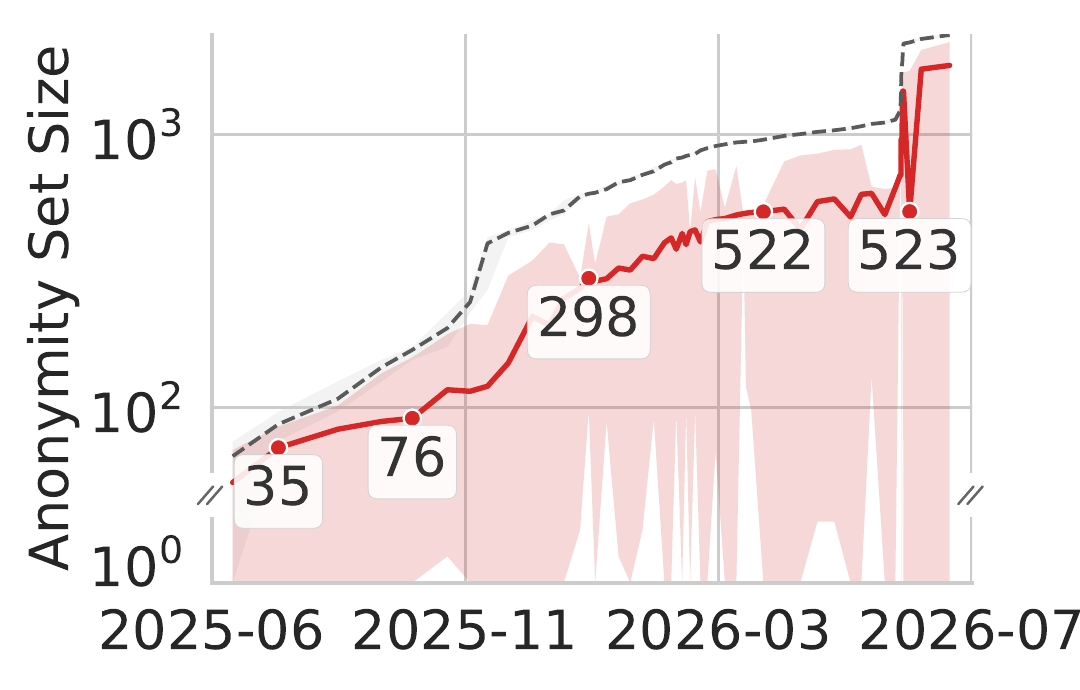}
        \caption{Base}
    \end{subfigure}\hspace{0.001\textwidth}%
    \begin{subfigure}[t]{0.199\textwidth}
        \centering
        \includegraphics[width=\linewidth, trim=0 10 0 0, clip]{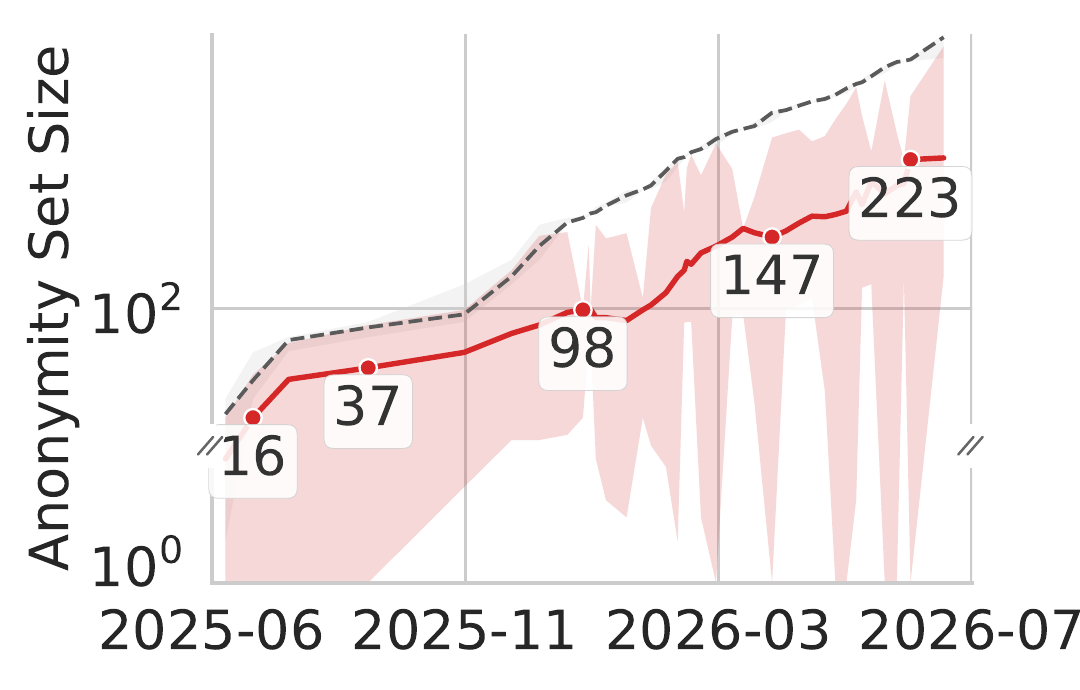}
        \caption{Optimism}
    \end{subfigure}

    \par\vspace{0.4em}
    \caption{Hinkal Anonymity Set Size over time: temporal baseline (dashed) and final transaction-level union after closure lifting (solid). P50 curves use equal-count block-order bins of about 100 transactions; P0--P100 shading and labels at the 5\%, 25\%, 50\%, 75\%, and 95\% timeline positions.}
    \label{fig:hinkal_anonymity_set_size_multichain}
\end{figure*}

\begin{table}[t]
    \centering
    \tiny
    \newcommand{\tabtypesymA}[1]{\raisebox{-0.18ex}{\scalebox{0.72}{\fcirc{#1}}}}
    \setlength{\tabcolsep}{1.8pt}
    \renewcommand{\arraystretch}{1.08}
    \caption{Anonymity Set Size reduction by proof composition.}
    \label{tab:as_summary}

    \begin{subtable}{\linewidth}
        \caption{Railgun.}
        \label{tab:railgun_as_summary}
        \centering
        \resizebox{\linewidth}{!}{%
            \begin{tabular}{lcc cc cc cc}
                \toprule
                \multirow{2}{*}{Class}                  & \multicolumn{2}{c}{\textbf{Ethereum}} & \multicolumn{2}{c}{\textbf{Arbitrum}} & \multicolumn{2}{c}{\textbf{Polygon}} & \multicolumn{2}{c}{\textbf{BNB Chain}}                                         \\
                \cmidrule(lr){2-3}\cmidrule(lr){4-5}\cmidrule(lr){6-7}\cmidrule(lr){8-9}
                                                        & Count                                 & Mean                                  & Count                                & Mean                                   & Count    & Mean   & Count    & Mean   \\
                \midrule
                All                                     & 85{,}157                              & 40.1\%                                & 42{,}256                             & 46.3\%                                 & 27{,}719 & 59.0\% & 14{,}708 & 47.6\% \\
                \midrule
                Proof \tabtypesymA{2}-only              & 10{,}025                              & 53.4\%                                & 6{,}611                              & 65.6\%                                 & 3{,}828  & 74.1\% & 1{,}992  & 58.1\% \\
                \quad 1 proof                           & 8{,}942                               & 55.8\%                                & 6{,}392                              & 66.4\%                                 & 3{,}689  & 74.9\% & 1{,}853  & 60.0\% \\
                \quad $>1$ proofs                       & 1{,}083                               & 33.6\%                                & 219                                  & 44.4\%                                 & 139      & 53.8\% & 139      & 32.8\% \\
                \midrule
                Proof \tabtypesymA{4}-only              & 74{,}352                              & 38.4\%                                & 35{,}331                             & 42.8\%                                 & 23{,}574 & 56.7\% & 12{,}434 & 46.2\% \\
                \quad 1 proof                           & 73{,}967                              & 38.4\%                                & 35{,}114                             & 42.9\%                                 & 23{,}244 & 56.9\% & 12{,}345 & 46.3\% \\
                \quad $>1$ proofs                       & 385                                   & 24.0\%                                & 217                                  & 24.6\%                                 & 330      & 40.1\% & 89       & 32.0\% \\
                \midrule
                Mixed (\tabtypesymA{2}+\tabtypesymA{4}) & 780                                   & 32.6\%                                & 314                                  & 31.9\%                                 & 317      & 46.2\% & 282      & 35.3\% \\
                \bottomrule
            \end{tabular}%
        }
    \end{subtable}

    \vspace{5pt}

    \begin{subtable}{\linewidth}
        \caption{Hinkal.}
        \label{tab:hinkal_as_summary}
        \centering
        \resizebox{\linewidth}{!}{%
            \begin{tabular}{lcc cc cc cc cc}
                \toprule
                \multirow{2}{*}{Class}                  & \multicolumn{2}{c}{\textbf{Ethereum}} & \multicolumn{2}{c}{\textbf{Arbitrum}} & \multicolumn{2}{c}{\textbf{Polygon}} & \multicolumn{2}{c}{\textbf{Base}} & \multicolumn{2}{c}{\textbf{Optimism}}                                                \\
                \cmidrule(lr){2-3}\cmidrule(lr){4-5}\cmidrule(lr){6-7}\cmidrule(lr){8-9}\cmidrule(lr){10-11}
                                                        & Count                                 & Mean                                  & Count                                & Mean                              & Count                                 & Mean   & Count   & Mean   & Count   & Mean   \\
                \midrule
                All                                     & 1{,}884                               & 57.0\%                                & 1{,}613                              & 57.4\%                            & 3{,}156                               & 53.0\% & 5{,}851 & 52.0\% & 4{,}012 & 48.9\% \\
                \midrule
                Proof \tabtypesymA{2}-only              & 1{,}019                               & 56.2\%                                & 767                                  & 59.1\%                            & 1{,}568                               & 53.5\% & 3{,}622 & 50.5\% & 2{,}291 & 48.1\% \\
                \quad 1 proof                           & 992                                   & 56.9\%                                & 739                                  & 59.9\%                            & 1{,}464                               & 54.9\% & 3{,}429 & 51.0\% & 2{,}149 & 49.4\% \\
                \quad $>1$ proofs                       & 27                                    & 29.2\%                                & 28                                   & 40.2\%                            & 104                                   & 34.8\% & 193     & 41.7\% & 142     & 27.7\% \\
                \midrule
                Proof \tabtypesymA{4}-only              & 848                                   & 58.3\%                                & 835                                  & 56.2\%                            & 1{,}566                               & 52.6\% & 2{,}178 & 54.6\% & 1{,}695 & 50.4\% \\
                \quad 1 proof                           & 821                                   & 59.5\%                                & 813                                  & 56.9\%                            & 1{,}505                               & 53.7\% & 2{,}100 & 55.2\% & 1{,}609 & 51.9\% \\
                \quad $>1$ proofs                       & 27                                    & 22.7\%                                & 22                                   & 27.5\%                            & 61                                    & 26.2\% & 78      & 38.1\% & 86      & 23.1\% \\
                \midrule
                Mixed (\tabtypesymA{2}+\tabtypesymA{4}) & 17                                    & 34.9\%                                & 11                                   & 29.2\%                            & 22                                    & 44.0\% & 51      & 45.0\% & 26      & 29.0\% \\
                \bottomrule
            \end{tabular}%
        }
    \end{subtable}
\end{table}

\begin{table}[t]
\centering
\tiny
\setlength{\tabcolsep}{2.0pt}
\renewcommand{\arraystretch}{1.08}
\caption{Anonymity Set Size reduction by depth quartile.}
\label{tab:closure_depth_summary}
\begin{subtable}{\linewidth}
\caption{Railgun.}
\label{tab:railgun_closure_depth_summary}
\centering
\resizebox{\linewidth}{!}{%
\begin{tabular}{@{}lrrrrrrr@{}}
\toprule
Chain & Transactions & One contributing proof (\%) & Median max. hops & Q1 & Q2 & Q3 & Q4 \\
\midrule
Ethereum & 85{,}157 & 82{,}909 (97.4\%) & 24{,}872 & 54.5\% & 51.1\% & 43.3\% & 12.3\% \\
Arbitrum & 42{,}256 & 41{,}506 (98.2\%) & 12{,}005 & 59.8\% & 57.2\% & 52.0\% & 17.1\% \\
Polygon & 27{,}719 & 26{,}933 (97.2\%) & 7{,}727 & 66.0\% & 64.0\% & 64.0\% & 43.6\% \\
BNB Chain & 14{,}708 & 14{,}198 (96.5\%) & 4{,}628.5 & 59.8\% & 49.5\% & 43.6\% & 39.7\% \\
\bottomrule
\end{tabular}%
}
\end{subtable}
\vspace{5pt}
\begin{subtable}{\linewidth}
\caption{Hinkal.}
\label{tab:hinkal_closure_depth_summary}
\centering
\resizebox{\linewidth}{!}{%
\begin{tabular}{@{}lrrrrrrr@{}}
\toprule
Chain & Transactions & One contributing proof (\%) & Median max. hops & Q1 & Q2 & Q3 & Q4 \\
\midrule
Ethereum & 1{,}884 & 1{,}813 (96.2\%) & 645 & 69.9\% & 57.8\% & 57.1\% & 47.5\% \\
Arbitrum & 1{,}613 & 1{,}552 (96.2\%) & 821 & 64.9\% & 56.5\% & 56.6\% & 55.3\% \\
Polygon & 3{,}156 & 2{,}969 (94.1\%) & 1{,}531.5 & 64.7\% & 57.6\% & 49.1\% & 45.6\% \\
Base & 5{,}851 & 5{,}529 (94.5\%) & 2{,}608 & 73.5\% & 45.8\% & 40.3\% & 50.8\% \\
Optimism & 4{,}012 & 3{,}758 (93.7\%) & 1{,}697.5 & 61.6\% & 44.9\% & 44.6\% & 50.7\% \\
\bottomrule
\end{tabular}%
}
\end{subtable}
\end{table}

\subsection{Supporting Commitment-Set Analyses}
\label{app:supporting_commitment}
\label{app:token-usd-aggregation}

This appendix supplies the Hinkal time-evolution view and the pricing coverage for the descriptive Railgun token-activity insets in \autoref{ssec:results_omega}.

\point{Hinkal Commitment Set Size}
\autoref{fig:hinkal_commitment_set_size_multichain} complements the Railgun chronology in \autoref{fig:commitment_set_size_multichain}. Each selected Hinkal pool has one long-lived tree and no Railgun-style tree number, so its panels compare the temporal baseline with Proof Root, Token, and the propagated final set after Value. They use the same cumulative sets and block-order aggregation as \autoref{ssec:results_omega}.

\point{USD Aggregation for Token-Activity Insets}
The inset pies of \autoref{fig:top_tokens_activity_multichain} describe value concentration for the token categories in the count-based bars; they do not affect pruning or closure lifting.

We count shield-side value once per commitment with a public plaintext amount and unshield-side value once per public Unshield record. One record may correspond to several nullifiers but exposes only one released amount, so that amount is never multiplied by its nullifier count. Tokens are matched by chain and contract address; \href{https://coinmarketcap.com/}{CoinMarketCap} and \href{https://www.coingecko.com/}{CoinGecko} provide metadata and spot prices. A fungible token contributes only when this exact-address lookup resolves a USD price. The pies use the same top-five categories as the bars and merge the remainder into Others.

Exact-address prices cover 99.17\%, 99.26\%, 97.66\%, and 97.55\% of ERC-20 public-value records on Ethereum, Arbitrum, Polygon, and BNB Chain, respectively. This denominator differs from the unshield bars, which count consumed nullifiers. The bars therefore cover all decoded activity, whereas the USD-value pies omit unpriced long-tail tokens and \acs{NFT}s.

\subsection{Supporting Anonymity-Set Analyses}
\label{app:supporting_anonymity}

This appendix provides the Hinkal Anonymity Set Size chronology and the chain-level proof-composition and closure-depth results supporting \autoref{ssec:results_as}.

\point{Hinkal Anonymity Set Size}
\autoref{fig:hinkal_anonymity_set_size_multichain} complements the \mbox{Railgun} chronology in \autoref{fig:anonymity_set_size_multichain}, using the same temporal baseline and final transaction-level Anonymity Set after closure lifting and within-transaction union.

\point{Token Partition Through Closure}
To test whether Hinkal's public token information remains constraining after closure lifting, we classify an Ethereum unshielding spend transaction as token-pure when it publicly releases one token and every feasible terminal shielding source reached by the complete closure has that same token. Of the 1{,}884 evaluated transactions, 1{,}813 (96.2\%) are token-pure, none of the single-token transactions has cross-token ancestry, and the remaining 71 publicly release multiple tokens. Within the token-pure cohort, the mean temporal shielding-address baseline is 484.98; restricting that baseline to the public token leaves 197.73 addresses, and the final Anonymity Set after complete closure has a mean size of 197.47. Thus, the public token partition persists from commitment candidates to address-level provenance. Unlike the one-contributor stratification in \autoref{tab:closure_depth_summary}, token ancestry classifies terminal-source tokens rather than proof sets unioned at the transaction layer.

\point{Proof-Composition Breakdown}
\autoref{tab:as_summary} partitions each chain's unshielding transactions into Proof~\fcirc{2}-only, Proof~\fcirc{4}-only, and mixed classes. Count gives the number of transactions in each class, and Mean is its average Anonymity Set Size reduction relative to $\mathcal{A}_0(T)$. Indented rows distinguish whether one or multiple proofs contribute to $\mathcal{A}(T)$, exposing how proof form interacts with within-transaction union; the strata are descriptive rather than causal.

\point{Closure-Depth Analysis}
One closure hop crosses one historical private state transition. For each transaction, maximum depth is the longest feasible path from a contributing proof to a Shield source. To separate depth from within-transaction union, we rank only transactions for which exactly one proof contributes to $\mathcal{A}(T)$, independently within each chain, and divide the complete ordering into four near-equal rank quartiles from shallowest (Q1) to deepest (Q4).

\autoref{tab:closure_depth_summary} covers all 169{,}840 Railgun and 16{,}516 Hinkal unshielding transactions, each with a matching depth record. The one-contributor comparison retains 96.5\%--98.2\% and 93.7\%--96.2\% of the respective chain cohorts. Median hops use all transactions; Q1--Q4 report mean Anonymity Set Size reduction within the ranked one-contributor cohort. Reduction declines from Q1 to Q4 on every Railgun chain, and Hinkal likewise has lower Q4 than Q1 on every chain. Base and Optimism nevertheless rebound in Q4 relative to Q3, showing that depth is informative but not sufficient by itself: token composition and pool history also shape the final set. These descriptive associations do not isolate depth from growth in protocol history and its temporal baseline.

\endgroup

\end{document}